\documentclass[review,3p,times]{elsarticle}
\usepackage{amsmath,hyperref}
\usepackage{lineno}
\usepackage{color}
\usepackage{subfigure}
\usepackage{epsfig}
\usepackage{caption,xcolor,float,graphicx}
\usepackage{graphics}
\usepackage{epstopdf}
\journal{Physical Review E}

\begin{document}

\begin{frontmatter}
	
\title{Phase-field-based lattice Boltzmann model for simulating thermocapillary flows}
\author[mymainaddress]{Lei Wang\corref{mycorrespondingauthor}}
\cortext[mycorrespondingauthor]{Corresponding author}
\ead{wangleir1989@126.com}
\author[mymainaddress]{Kun He}
\author[mysecondaddress]{Huili Wang}
\address[mymainaddress]{School of Mathematics and Physics, China University of Geosciences, Wuhan 430074, China}
\address[mysecondaddress]{3School of Mathematical and Computer Sciences, Wuhan Textile University, Wuhan 430200, China}

\begin{abstract}
This paper proposes a simple and accurate lattice Boltzmann model for simulating thermocapillary flows, which is able to deal with thermophysical parameters contrasts. In this model, two lattice Boltzmann equations are utilized to solve the conservative Allen-Cahn equation and the incompressible Navier-Stokes equations, while another lattice Boltzmann equation is used for solving the temperature field, where the collision term is delicately designed such that the influence of the thermophysical parameters contrasts is incorporated. In contrast to previous lattice Boltzmann model for thermocapillary flows, the most distinct feature of the current model is that the forcing term used in the present thermal lattice Boltzmann equation is not needed to calculate space derivatives of the heat capacitance or the order parameter, making the scheme much simpler and also possible to retain the main merits of the lattice Boltzmann method. The developed model is firstly validated by  considering the thermocapillary flows in a heated microchannel with two superimposed planar fluids. It is then used to simulate thermocapillary migration of a two-dimensional deformable droplet, and its accuracy is once again consistent with the theoretical prediction when the Marangoni number approaches zero.  Finally, we numerically study the motion of two recalcitrant bubbles in a two-dimensional channel where the relationship between surface tension and temperature is assumed to be a parabolic function. It is found that owing to the competing between the inertia and thermal effects, the bubbles are able to move against the liquid's bulk motion and towards areas with low surface tension. 

\end{abstract}
	
\begin{keyword}
Thermocapillary flows\sep Lattice Boltzmann method \sep Thermophysical parameters contrasts
\end{keyword}	 
\end{frontmatter}

\section{Introduction}
Multiphase fluid interactions are nearly ubiquitous in natural and industrial processes, and such flows have attracted considerable attention for decades. As is well known, an interaction between two different fluid molecules across the phase interface will induce a surface tension, which  is a state function, and its value usually depends on the temperature, composition, and electric charge density of the fluids \cite{striven1960}. Particularly, when the temperature distribution in a multiphase system is nonuniform, the temperature gradient along the interface will invoke a surface tension gradient, resulting in the fluids moving from the regions of high surface tension to the areas of low surface tension \cite{scriven1960nature}. The fluid motion in these phenomenons are commonly called thermocapillary flows or thermal Marangoni flows, which play important roles in several systems, such as crystal-growth processes \cite{shen1990jfm}, welding pools \cite{mills1998marabgo}, droplet coalescence and formation \cite{jasnow1996coarse}, etc, and consequently attracts considerable scientific interest in the last few years.

Historically, the study on thermocapillary flows dates back to the pioneering work of Young et al. in 1959 \cite{young1959the}, in which they came up with an analytical solution for the terminal velocity of a spherical droplet in an infinite ambient fluid. Since then, many experimental attempts have been conducted to reveal the primary mechanism of thermocapillary flows \cite{schatz2001experiment,robert2012thermal,kamotani2000micro,kang2019the}.  However,  because of the small spatial and temporal scales involved in such flows, performing precise experimental measurements of the local temperature and velocity fields is still a challenging task \cite{karbalaei2016thermo}. As another effective way, numerical simulation is thus of great significance  to study the thermohydrodynamics aspects of the thermocapillary flows in detail. To date, a number of continuum methods on a macroscopic scale have been proposed to deal with the thermocapillary flows, among which the front-tracking method \cite{nas2033therm}, the volume-of-fluid (VOF) method \cite{ma2013numerical} and the level-set (LS) method \cite{zhao2011topo} are commonly used. However, it has been reported that the front-tracking method is not suitable for simulating interface breakup and coalescence, and the hyperbolic character of interface equation in VOF and LS methods sets high demand on the numerical stability \cite{shyy1996comput}. Actually, it is well understood that the interfacial phenomenon is  a natural consequence of microscopic interactions among fluid molecules. In this setting, the lattice Boltzmann (LB) method, which is built upon the molecular kinetic theory \cite{kruge2017the,guobook2013lattice}, is a suitable candidate to describe such interactions process at microscopic level.

Over the past decades, the LB method has become a numerically robust and efficient technique for simulating  multiphase fluids \cite{huang2015multiphase,lireview2016lattice,huang2022three,wang2022thermal}.  Compared with conventional methods for multiphase flows, the multiphase models employed in LB community usually belong to the diffusive interface approach, in which the physical parameter profile smoothly varies across the interface between the two bulk values. With this feature, the motion of the interface need not be tracked explicitly \cite{huang2015multiphase,lireview2016lattice}. 
Thus, some extra efforts such as interface reconstruction or reinitialization encountered in conventional methods are avoided. In the context of LB method, several types of thermocapillary LB models have been developed from different perspectives of the interactions, which commonly include the pseudopotential-based model, color-gradient-based model, and  phase-field-based model.   Gupta et al. developed a pseudopotential-based thermocapillary model to simulate the effects of thermocapillary on droplet formation in a microfluidic T-junction \cite{gupta2016lattice}. However, it is known that the surface tension, which is a key parameter in thermocapillary flows and is usually a function of temperature, in the pseudopotential LB model cannot be prescribed in advance, but instead is determined by a series of prior simulations \cite{lireview2016lattice}. It thus limits its prospect for practical applications.   Liu et al. also proposed a color-gradient-based LB model to simulate thermocapillary flows \cite{liu2012modeling}, in which the phase interface behavior is captured by using the color-gradient model, and the temperature field is solved by another LB equation. With this model, the authors successfully simulated the thermocapillary flow for droplet manipulation in a microchannel. Later, Liu and the cooperators further extended the proposed model to model immiscible thermocapillary flows with the presence of fluid-surface interactions \cite{liujcp2015modelling} and axisymmetric thermocapillary flows \cite{liu2017alattice}. Nevertheless, it has the deficiency that these models are only able to handle thermocapillary flows with equal or small density contrasts.  In addition,  because the temperature equation that appeared in these models is a simplified one, the specific heat capacities in both fluids are assumed to be the same. Particularly, it is noticed that this assumption also exists in the recent works by Liu et al. \cite{liujcp2014lattice} and Zheng et al. \cite{zheng2016continuous} which suggests that the above models will suffer from some limitations more or less in practical applications. To remove these limitations, Liu et al. \cite{liupre2013phase} developed a phase-field-based hybrid LB model for thermocapillary flows, where the interface equation described by the Cahn-Hilliard equation is solved with Lee et al.'s LB model \cite{leejcp2010lattice}, and the temperature field is solved through a combination of the finite-difference method and the fourth-order Runge–Kutta method.  Although their model is able to simulate thermocapillary flows with relatively large density ratios and thermophysical parameters contrasts,  the discretization schemes used for interfacial forces usually introduce incompatible numerical errors, leading to the violation of mass and momentum conservation \cite{lou2012effects}. Besides, while the finite difference method is simple in principle, it is not suitable for the study of thermocapillary flows with complex boundaries, meaning that it is unable to retain the original merit of the LB method. Note that the above hybrid scheme is still the widely used approach in simulation of thermocapillary flows \cite{qiaomodel2018ate,majidi2020single,mitchell2021compu}, and the reason behind this phenomenon lies in that the temperature equation encountered in this case is not a standard convection-diffusion equation, leading it cannot be directly solved with the conventional thermal LB approach \cite{wang2019ammalatt,guo2002acoupled}. To model thermocapillary flows in the framework of LB method, recently Hu et al. \cite{hu2019adiffuse}, and Yue et al. \cite{yue2022improved} developed two different phase-field-based LB models for thermocapillary flows. Compared with previous phase-field-based LB approaches, their models use the Allen-Cahn equation to capture the phase interface which shares better volume preservation property and numerical stability \cite{wang2016comparative}. However, since the thermal LB models in these two works are both constructed by reformulating the temperature equation as a standard convection-diffusion equation with a source term, their thermal models need to estimate the space derivatives of the heat capacitance or the order parameter, which increases the difficulty of implementation and may also affect the numerical stability of the thermal LB approach.

In this paper, we intend to the proposed a unified phase-field-based LB model for simulating thermocapillary flows, in which the hydrodynamic equations i.e., the Navier-Stokes and the Allen-Cahn equations are solved with the improved LB model proposed by Liang et al. \cite{liang2018phase}, while the temperature equation is solved by another elaborately designed LB scheme. Particularly, since the present thermal model needs not to calculate the gradients of the heat capacitance or the order parameter, it is much simpler than previous models. Also, the present model is able to simulate thermocapillary flowsI with different heat capacitances or thermal conductivities. The reminder of the present paper is organized as follows. Section 2 briefly introduces the phase-field theory and conservation equations. Section 3 gives the proposed LB method for thermocapillary flows.  After that, three numerical experiments are present in  Section 4, and a brief conclusion is drawn in Section 5.

\section{Phase-field theory and conservation equations}
The phase-field method is one of the most popular diffusive interface approaches for modeling multiphase flows. In this method, the sharp interface among different phases is replaced by thin but nonzero thickness transition regions, where the fluid variables smoothly vary across the interface \cite{huang2015multiphase}. To distinguish the different fluid phases in a multiphase flow, the phase-field method needs another interface-capturing equation to describe the evolution of the phase-field parameter (also called the order parameter) \cite{kruge2017the}. Currently, the most applied interface-capturing equation in phase-field modeling is the so-called Cahn-Hilliard equation since it is able to ensure the mass conservation of multiphase system \cite{huang2015multiphase}. However, as pointed out by some researchers,  owing to the Cahn-Hilliard equation is actually a fourth-order partial differential equation, the collision process of the corresponding LB models for Cahn-Hilliard equation cannot be performed locally \cite{wang2016comparative,liang2018phase}. To address this issue, this paper adopts the second-order conservative Allen-Cahn equation to track the multiphase interface, which has been reported more accurate and stable than the Cahn-Hilliard equation.

To facilitate the discussion, in what follows we briefly introduce the Allen-Cahn equation with mass conservation, and a more detailed description on this equation can be found in the recent review of \cite{wang2019abrief}. According to the work of Sun and Beckermann \cite{sun2007jcpsharp}, it is known that the interface advection equation of the order parameter can be given by 
\begin{equation}
\frac{{\partial \phi }}{{\partial t}} + \left( {{u_n}{\bf{n}} + {\bf{u}}} \right) \cdot \nabla \phi  = 0,\label{eq_1}
\end{equation}
where $\phi$ is the order parameter,  ${\bf{u}}$ and  ${\bf{n}} = {{\nabla \phi } \mathord{\left/{\vphantom {{\nabla \phi } {\left| {\nabla \phi } \right|}}} \right.\kern-\nulldelimiterspace} {\left| {\nabla \phi } \right|}}$ in Eq. (\ref{eq_1}) are the fluid velocity and the unit normal vector, respectively. In addition,  ${u_n}$ is the normal interface speed defined as ${u_n} =  - {M_{\phi}}{\kappa _\phi }$, in which $M_{\phi}$ is the mobility and ${\kappa _\phi }$  is the interface curvature given by 
\begin{equation}
{\kappa _\phi } = \nabla  \cdot {\bf{n}} = \frac{1}{{\left| {\nabla \phi } \right|}}\left[ {{\nabla ^2}\phi  - \frac{{\left( {\nabla \phi  \cdot \nabla } \right)\left| {\nabla \phi } \right|}}{{\left| {\nabla \phi } \right|}}} \right].	
\label{eq_2}
\end{equation}
In this work we use ${\phi _l} =  - 1$ to denote the light fluid, while the heavy fluid is identified by ${\phi _h} =  1$. In this setting, the interface between different fluids can be described by ${\phi} =  0$ with an interfacial layer of thickness of $W$. In particular, the order profile at the equilibrium state can be written as
\begin{equation}
\phi \left( {\bf{x}} \right) = 0.5\tanh \left( {2\frac{{\left\| {{\bf{x}} - {{\bf{x}}_0}}, \right\|}}{W}} \right)	
\label{eq_3}
\end{equation}
in which ${\left\| {{\bf{x}} - {{\bf{x}}_0}} \right\|}$ denotes the spatial location normal to the interface ${{{\bf{x}}_0}}$.  Following the idea of Folch et al. \cite{folch1999phase}, a counter term is introduced into Eq. (\ref{eq_2}) to handle the case of no curvature-driven interface motion, and thus we have, 
\begin{equation}
\frac{{\partial \phi }}{{\partial t}} + {\bf{u}} \cdot \nabla \phi  = {M_{\phi}}\left[ {{\nabla ^2}\phi  - \frac{{\left( {\nabla \phi  \cdot \nabla } \right)\left| {\nabla \phi } \right|}}{{\left| {\nabla \phi } \right|}} - \left| {\nabla \phi } \right|\nabla  \cdot \left( {\frac{{\nabla \phi }}{{\left| {\nabla \phi } \right|}}} \right)} \right].
\label{eq_4}
\end{equation}
With the assumption of the incompressible condition, i.e., $\nabla  \cdot {\bf{u}} = 0$, and note that $\left| {\nabla \phi } \right| =  - {{\phi \left( {1 - {\phi ^2}} \right)} \mathord{\left/
{\vphantom {{\phi \left( {1 - {\phi ^2}} \right)} {{W^2}}}} \right.\kern-\nulldelimiterspace} {{W^2}}}$ \cite{wang2019abrief},  the second-order local conservative Allen-Cahn equation is finally expressed as 
\begin{equation}
\frac{{\partial \phi }}{{\partial t}}+\nabla \cdot(\phi \mathbf{u})=\nabla \cdot M_{\phi} \left( \nabla \phi-\frac{\nabla \phi}{|\nabla \phi|} \frac{1-\phi^{2}}{\sqrt{2} W} \right).
\label{eq_5}
\end{equation}

Apart from the above mentioned interface-capturing equation, we further intend to introduce the governing equations for two-phase flows. To this end, we first assume the two fluids are immiscible and Newtonian, and then the Navier-Stokes equations can be written as \cite{unverdi1992afront}   
\begin{equation}
\nabla  \cdot {\bf{u}} = 0,
\label{eq_6}	
\end{equation}
\begin{equation}
\rho \left( {\frac{{\partial {\bf{u}}}}{{\partial t}} + {\bf{u}} \cdot \nabla {\bf{u}}} \right) =  - \nabla p + \nabla  \cdot \left[ {\mu \left( {\nabla {\bf{u}} + \nabla {{\bf{u}}^{\rm T}}} \right)} \right] + {{\bf{F}}_s} + {\bf{G}},
\label{eq_7}
\end{equation}
where $\rho$ is the fluid density determined by 
\begin{equation}
	\rho=\frac{\rho _l-\rho_ h}{\phi_l - \phi_h}(\phi - \phi_h)+\rho_h,
	\label{eq_8}	
\end{equation}  
$p$ and $\mu$ represent the hydrodynamic pressure and fluid dynamics viscosity, respectively.  ${\bf{G}}$ is the additional body force, and ${{\bf{F}}_s}$ is the volume-distributed interfacial force, which can be written as \cite{wang2019abrief} 
\begin{equation}
{{\bf{F}}_s} = \left[ { - \sigma {\kappa _\phi }{\bf{n}} + \left( {{\bf{I}} - {\bf{n}} \otimes {\bf{n}}} \right) \cdot \nabla \sigma } \right]{\delta _\Sigma },
\label{eq_9}
\end{equation}
where $\sigma $ and ${\bf{I}}$ represent the local surface tension and the identity matrix, respectively. ${\delta _\Sigma }$ is the Dirac delta function satisfying
\begin{equation}
\int_{ - \infty }^{ + \infty } {{\delta _\Sigma }dx}  = 1, 
\label{eq_10}
\end{equation}
and in this work, it is defined as the form of ${\delta _\Sigma } = {{3\sqrt 2 W{{\left| {\nabla \phi } \right|}^2}} \mathord{\left/{\vphantom {{3\sqrt 2 W{{\left| {\nabla \phi } \right|}^2}} 4}} \right.\kern-\nulldelimiterspace} 4}$. In such a case, the total interfacial force can be simplified as 
\begin{equation}
\begin{array}{l}
		{{\bf{F}}_s}  = \frac{{3\sqrt 2 W}}{4}\nabla  \cdot \left[ {\sigma {{\left| {\nabla \phi } \right|}^2}{\bf{I}} - \sigma \nabla \phi \nabla \phi } \right]\\
		\;\;\;\;\;= \frac{{3\sqrt 2 W}}{4}\left[ {{{\left| {\nabla \phi } \right|}^2}\nabla \sigma  + \sigma \nabla {{\left| {\nabla \phi } \right|}^2} - \nabla \sigma  \cdot \left( {\nabla \phi \nabla \phi } \right) - \sigma \nabla  \cdot \left( {\nabla \phi \nabla \phi } \right)} \right]\\
	\;\;\;\;\;	= \frac{{3\sqrt 2 W}}{4}\left[ {{{\left| {\nabla \phi } \right|}^2}\nabla \sigma  + \frac{1}{2}\sigma \nabla {{\left| {\nabla \phi } \right|}^2} - \nabla \sigma  \cdot \left( {\nabla \phi \nabla \phi } \right) - \sigma {\nabla ^2}\phi \nabla \phi } \right]\\
\;	\;\;\;\;	= \frac{{3\sqrt 2 W}}{4}\left[ {{{\left| {\nabla \phi } \right|}^2}\nabla \sigma  - \nabla \sigma  \cdot \left( {\nabla \phi \nabla \phi } \right) + \frac{\sigma }{{{W^2}}}{\mu _\phi }\nabla \phi } \right],
\end{array}
\label{eq_11}
\end{equation} 
where ${{\mu _\phi }}$ is the chemical potential, and it can be derived based on the free energy of double-well form as ${\mu _\phi } = \left( {\phi  - {\phi _l}} \right)\left( {\phi  - {\phi _h}} \right)\left[ {\phi  - 0.5\left( {{\phi _l} + {\phi _h}} \right)} \right] - {W^2}{\nabla ^2}\phi$ \cite{wang2019abrief}.   
To incorporate the thermocapillary effects, one must define an equation of state to relate the surface tension  $\sigma$  and temperature $T$. In this work we will focus on a linear relation between  $\sigma$  and $T$ unless mentioned otherwise, and it is defined as \cite{ma2011direct}
\begin{equation}
\sigma \left( T \right) = \sigma \left( {{T_{ref}}} \right) + {\sigma _T}\left( {T - {T_{ref}}} \right),
\end{equation}
where ${{T_{ref}}}$ is the reference temperature, and  ${\sigma _T} = {{\partial \sigma \left( T \right)} \mathord{\left/{\vphantom {{\partial \sigma \left( T \right)} {\partial T}}} \right.
\kern-\nulldelimiterspace} {\partial T}}$ is the rate of variation of the surface tension with the temperature.    

In a thermocapillary flow, an important aspect in understanding the thermal multiphase flows is to obtain  the time evolution of the temperature field. Physically, the conservation of energy in the form of a temperature equation can be expressed as \cite{ma2013jcpnumeri} 
\begin{equation}
\rho {c_p}\frac{{DT}}{{Dt}} = \nabla  \cdot \lambda \nabla T + \nabla {\bf{u}}:{\bf{S}} + \rho T\left( {\frac{\partial }{{\partial T}}\frac{1}{\rho }} \right)\frac{{Dp}}{{Dt}},
\end{equation}
where $\lambda$ is the thermal conductivity, and $c_p$ is the specific heat capacity, ${D \mathord{\left/
{\vphantom {D {Dt}}} \right.\kern-\nulldelimiterspace} {Dt}} = {\partial _t} + {\bf{u}} \cdot \nabla $ is the material derivative. ${\bf{S}} = \mu \left[ {\nabla {\bf{u}} + {{\left( {\nabla {\bf{u}}} \right)}^{\rm T}}} \right]$ is the viscous stress tensor. Note that the term $\nabla {\bf{u}}:{\bf{S}}$ denotes the viscous heating, which is neglected in this work. Also, since the fluid in the bulk is assumed to be incompressible, the last term is also dropped. In this setting, the governing equation for temperature field can be given by  
\begin{equation}
\rho {c_p}\left( {\frac{{\partial T}}{{\partial t}} + {\bf{u}} \cdot \nabla T} \right) = \nabla  \cdot \lambda \nabla T, 
\label{eq_13}	
\end{equation}
where $\lambda$ is the thermal conductivity, and $c_p$ is the specific heat capacity.
 
\section{Lattice Boltzmann method for thermocapillary flows}

In terms of the collision operator of the Boltzmann equation, the commonly used LB models can be classified into three kinds, i.e., the single-relaxation-time LB model (also called BGK model) \cite{qian1992lattice}, the two-relaxation-time LB model \cite{ginzburg2008two}, and the multiple-relaxation-time LB model \cite{lallemand2000theory}.  In this work, we intend to adopt the BGK model for its simplicity and efficiency, and it is not very difficult to extend the following BGK models to the other two collision operators. Besides, the present LB method for thermocapillary flows includes three LB equations, in which two LB equations are utilized to simulate the velocity field and the phase field, while another one is used for solving the temperature field. Particularly, the LB models for velocity and phase fields have been reported elsewhere, and here our focus is to develop a simple and improved LB model for the temperature field. 
\subsection{Lattice Boltzmann method for hydrodynamic equations}
We now turn to give the LB models for hydrodynamic equations, i.e., the Allen-Cahn equation [Eq. (\ref{eq_5})] and the Navier-Stokes equations [Eq. (\ref{eq_6}) and Eq. (\ref{eq_7})]. Based on  previous work, the LB equation for the Allen-Cahn equation can be written as  \cite{wang2016comparative}
\begin{equation}
\begin{aligned}
	f_{i}\left(\mathbf{x}+\mathbf{c}_{i} \Delta t, t+\Delta t\right)=f_{i}(\mathbf{x}, t)-\frac{1}{\tau_{\phi}}\left[f_{i}(\mathbf{x}, t)-f_{i}^{\mathrm{eq}}(\mathbf{x}, t)\right] 
	+\Delta t\left(1-\frac{1}{2 \tau_{\phi}}\right) R_{i}(\mathbf{x}, t),
\end{aligned}
\label{eq_14}
\end{equation}
where $f_{i}(\mathbf{x}, t)$ is the discrete distribution function of the order parameter $\phi$ 
at position $\mathbf{x}$ and time $t$, $f_i^{\mathrm{eq}}$ is the local equilibrium distribution function defined by \cite{wang2016comparative} 
\begin{equation}
	f_i^{\mathrm{eq}}(\mathbf{x}, t)=\omega_{i}\phi\left(1+\frac{\mathbf{c}_i\cdot \mathbf{u}}{c_s^{2}}\right).		
\end{equation}
Here, $\omega_{i}$ and ${c_s}$ represent the weight coefficient and the lattice sound speed, and their values depend on the discrete velocity set $\mathbf{c}_i$ which will be given later. $R_{i}(\mathbf{x}, t)$ is the discrete source term designed as  \cite{wang2016comparative}
\begin{equation}
	\begin{aligned}
		R_{i}(\mathbf{x}, t)=\omega_{i}\frac{ \mathbf{c}_{i} \cdot\left[\partial_{t}(\phi \mathbf{u})+c_{s}^{2}\frac{\nabla \phi}{|\nabla \phi|} \frac{1-\phi^{2}}{\sqrt{2} W}\right]}{c_{s}^{2}}.
	\end{aligned}
\end{equation}    
By using the Chapman-Enskog analysis, the LB equation given by Eq. (\ref{eq_14}) can be recovered to the Allen-Cahn equation with the mobility given be the relaxation time $\tau_{\phi}$ as 
\begin{equation}
	M_{\phi}=c_s^2(\tau_{\phi}-0.5)\Delta t.
\end{equation}
In such a case, the order parameter can be calculated from $f_i$ as $\phi=\sum_{i}f_{i}$. Note that although there is a gradient term of ${\nabla \phi }$ in the source term $R_i$, it can be calculated from the local values of $f_i$ by  \cite{wang2016comparative} 
\begin{equation}
\nabla \phi  = \frac{{\sum\limits_i {{{\bf{c}}_i}\left( {{f_i} - f_i^{{\rm{eq}}}} \right)}  + 0.5\Delta t{\partial _t}\left( {\phi {\bf{u}}} \right)}}{{ - c_s^2{\tau _\phi }\Delta t + \frac{{{M_\phi }\left( {1 - 4{\phi ^2}} \right)}}{{W\left| {\nabla \phi } \right|}}}},
\end{equation}
where ${\left| {\nabla \phi } \right|}$ is derived with \cite{wang2016comparative}
\begin{equation}
\left| {\nabla \phi } \right| = \frac{{ - W\left| {\sum\limits_i {{{\bf{c}}_i}\left( {{f_i} - f_i^{{\rm{eq}}}} \right)}  + 0.5\Delta t{\partial _t}\left( {\phi {\bf{u}}} \right)} \right| - {M_\phi }\left( {1 - 4{\phi ^2}} \right)}}{{ - c_s^2{\tau _\phi }\Delta tW}}.	
\end{equation}

In addition to the LB equation for Allen-Cahn equation, we also need another LB model to solve the 
Navier-Stokes equations given by Eq. (\ref{eq_6}) and Eq. (\ref{eq_7}). Based on previous works, the collision-streaming process for Navier-Stokes equations reads  \cite{liang2018phase} 
\begin{equation}
	\begin{aligned}
		g_{i}\left(\mathbf{x}+\mathbf{c}_{i} \Delta t, t+\Delta t\right)=g_{i}(\mathbf{x}, t)-\frac{1}{\tau_{g}}\left[g_{i}(\mathbf{x}, t)-g_{i}^{\mathrm{eq}}(\mathbf{x}, t)\right] 
		+\Delta t\left(1-\frac{1}{2 \tau_{g}}\right) G_{i}(\mathbf{x}, t),
	\end{aligned}
\end{equation}
where $g_{i}(\mathbf{x}, t)$ is the discrete-velocity distribution function, $g_i^{{\rm{eq}}}$ is the equilibrium given by  \cite{liang2018phase}
\begin{equation}
	g_{i}^{\mathrm{eq}}= \begin{cases}\frac{p}{c_{s}^{2}}\left(\omega_{0}-1\right)+\rho s_{i}(\mathbf{u}), & i=0, \\ \frac{p}{c_{s}^{2}} \omega_{i}+\rho s_{i}(\mathbf{u}), & i \neq 0,\end{cases}
\end{equation}
with
\begin{equation}
	s_{i}(\mathbf{u})=\omega_{i}\left[\frac{\mathbf{c}_{i} \cdot \mathbf{u}}{c_{s}^{2}}+\frac{\left(\mathbf{c}_{i} \cdot \mathbf{u}\right)^{2}}{2 c_{s}^{4}}-\frac{\mathbf{u} \cdot \mathbf{u}}{2 c_{s}^{2}}\right].
\end{equation}
$ G_{i}(\mathbf{x}, t)$ is the discrete forcing term expressed as \cite{liang2018phase} 
\begin{equation}
{G_i} = {\omega _i}\left[ {{\bf{u}} \cdot \nabla \rho  + \frac{{{{\bf{c}}_i} \cdot \left( {{{\bf{F}}_s} + {\bf{G}}} \right)}}{{c_s^2}} + \frac{{{\bf{u}}\nabla \rho :\left( {{{\bf{c}}_i}{{\bf{c}}_i} - c_s^2{\bf{I}}} \right)}}{{c_s^2}}} \right].
\end{equation}
With the help of the Chapman-Enskog analysis again, the incompressible Navier-Stokes equations can be recovered up to second-order accurate with 
\begin{equation}
\mu  = \rho c_s^2\left( {{\tau _g} - \frac{1}{2}} \right)\Delta t,
\end{equation}
and the macroscopic variables (momentum and pressure) are then computed from 
\begin{equation}
	\rho \mathbf{u} =\sum_{i} \mathbf{c}_{i} g_{i}+0.5 \Delta t (\mathbf{F}_s + \bf{G}) ,
\end{equation}
\begin{equation}
	p =\frac{c_{s}^{2}}{\left(1-\omega_{0}\right)}\left[\sum_{i \neq 0} g_{i}+\frac{\Delta t}{2} \mathbf{u} \cdot \nabla \rho+\rho s_{0}(\mathbf{u})\right].
\end{equation}

%

\subsection{Modified lattice Boltzmann method for temperature equation}
In this subsection, we will proposed a modified LB method for solving the temperature equation in a thermocapillary flows. In contrast to previous works, the distinct features of the present LB model
lie in two aspects. For one thing, the present model considers the influence of the heat capacitances $\rho c_p$ between two fluids, making it possible to simulate  thermocapillary flows with thermophysical parameters contrasts. For another, our LB model is constructed without any modifications to the original temperature equation such that it is much simpler than some previous LB it methods for thermocapillary flows. To make it clearer, we first rewrite the temperature equation into another form,
\begin{equation}
\rho c_p \frac{\partial T}{\partial t} =\nabla \cdot (\lambda \nabla T) {- \rho c_p {\bf{u}} \cdot \nabla T },	
\label{eq_27}
\end{equation}
where the incompressible condition $\nabla  \cdot {\bf{u}} = 0$ is used. From Eq. (\ref{eq_27}), it is obvious that the temperature macroscopic equation has changed to a pure diffusion equation with a corresponding source term of $\widehat F =  - \rho {c_p}{\bf{u}} \cdot \nabla T$. While generally speaking it is easy to develop a LB model for diffusion equation with a source term, how to treat the term of $\left( {\rho {c_p}} \right)$ in front of ${{\partial T} \mathord{\left/{\vphantom {{\partial T} {\partial t}}} \right.\kern-\nulldelimiterspace} {\partial t}}$ is another problem that must be carefully considered. Inspired by previous works \cite{wang2022thermal,cartalade2016lattice, sun2019ananiso}, the LB equation with the single-relaxation-time collision operator for the temperature equations can be expressed as 
\begin{equation}
\rho {c_p}{h_i}\left( {{\bf{x}} + {{\bf{c}}_i}\Delta t,t + \Delta t} \right) - {h_i}\left( {{\bf{x}},t} \right) = \left( {\rho {c_p} - 1} \right){h_i}\left( {{\bf{x}} + {{\bf{c}}_i}\Delta t,t} \right) - \frac{1}{{{\tau _h}}}\left[ {{h_i}\left( {{\bf{x}},t} \right) - h_i^{eq}\left( {{\bf{x}},t} \right)} \right] + \Delta t{\widehat F_i}\left( {{\bf{x}},t} \right),
\label{eq_28}
\end{equation}
where ${h_i}\left( {{\bf{x}},t} \right)$ is the temperature distribution function, ${{\tau _h}}$ is the dimensionless relaxation time related to the conductivity, ${\widehat F_i}\left( {{\bf{x}},t} \right)$ is the discrete source term defined as 
\begin{equation}
{\widehat F_i}\left( {{\bf{x}},t} \right) = {\omega _i}\widehat F,
\label{weq_30}
\end{equation}
${h_i^{eq}\left( {{\bf{x}},t} \right)}$ is the equilibrium given by 
\begin{equation}
h_i^{eq}\left( {{\bf{x}},t} \right) = {\omega _i}T.
\label{weq_31}
\end{equation}
Note that in contrast to the commonly used collision step, there is a correction term $\left( {\rho {c_p} - 1} \right){h_i}\left( {{\bf{x}} + {{\bf{c}}_i}\Delta t,t} \right)$ appeared in the evolution equation, which is elaborately designed to incorporate the heat capacitances effects. Based on the above LB equation, the temperature is further defined as 
\begin{equation}
T = \sum\limits_i {{h_i} }.
\label{eq_31}
\end{equation}

To show the current LB model could recover the temperature equation precisely, we first apply the Taylor series expansion at second-order in space and first-order in time for Eq. (\ref{eq_28}), then one can get 
\begin{equation}
\rho {c_p}\left( {{h_i} + \Delta t{D_i}{h_i} + \frac{{\Delta {t^2}}}{2}\hat d_i^2{h_i}} \right) - {h_i} = \left( {\rho {c_p} - 1} \right)\left( {{h_i} + \Delta t{{\hat d}_i}{h_i} + \frac{{\Delta {t^2}}}{2}\hat d_i^2{h_i}} \right) - \frac{1}{{{\tau _h}}}\left( {{h_i} - h_i^{eq}} \right) + \Delta t{\widehat F_i},
\label{eq_32}
\end{equation}
where ${D_i} = {\partial _t} + {{\hat d}_i}$ with ${{\hat d}_i} = {{\bf{c}}_i} \cdot \nabla $. To deduce the temperature equation, the following Chapman-Enskog expansions are introduced
\begin{equation}
{\partial _t} = \sum\limits_{n = 1}^{ + \infty } {{\varepsilon ^n}{\partial _{tn}},} \;\;\;\;\; \nabla  = \varepsilon {\nabla _1},\;\;\;\;\;{h_i} = \sum\limits_{n = 1}^{ + \infty } {{\varepsilon ^n}h_i^{\left( n \right)},}\;\;\;\;\; \widehat F = \varepsilon {\widehat F^{\left( 1 \right)}},
\end{equation} 
where $\varepsilon$ is a small expansion parameter. Substituting the above expansions into Eq. (\ref{eq_32}), we can rewrite it in the consecutive orders of $\varepsilon$ as
\begin{flalign}
&{\varepsilon ^0}:\;\;\;\;\;h_i^{\left( 0 \right)} = h_i^{eq},&
\label{eq_34}
\end{flalign}  
\begin{flalign}
&{\varepsilon ^1}:\;\;\;\;\;\rho {c_p}{\partial _{t1}}h_i^{\left( 0 \right)} + {{\hat d}_{i1}}h_i^{\left( 0 \right)} =  - \frac{1}{{{\tau _h}\Delta t}}h_i^{\left( 1 \right)} + \widehat F_i^{\left( 1 \right)},&
\label{eq_35}
\end{flalign}
\begin{flalign}
&{\varepsilon ^2}:\;\;\;\;\;\rho {c_p}{\partial _{t1}}h_i^{\left( 1 \right)} + {{\hat d}_{i1}}h_i^{\left( 1 \right)} + \rho {c_p}{\partial _{t2}}h_i^{\left( 0 \right)} + \frac{{\Delta t}}{2}\hat d_{_{i1}}^2h_i^{\left( 0 \right)} =  - \frac{1}{{{\tau _h}\Delta t}}h_i^{\left( 2 \right)},&
\label{eq_36}
\end{flalign}
Eq. (\ref{eq_34}) suggests that 
\begin{equation}
\sum\limits_i {h_i^{\left( n \right)}}  = 0, \;\;\;\;\; \forall n \ge 1.
\end{equation}
Based on Eq. (\ref{weq_30}) and Eq. (\ref{weq_31}), we have 
\begin{equation}
\sum\limits_i {h_i^{\left( 0 \right)}}  = T,\;\;\;\;\;\sum\limits_i {{{\bf{c}}_i}h_i^{\left( 0 \right)}}  = 0,\;\;\;\;\;\sum\limits_i {{{\bf{c}}_i}{{\bf{c}}_i}h_i^{\left( 0 \right)}}  = c_s^2T,\;\;\;\;\;\sum\limits_i {\widehat F_i^{\left( 1 \right)}}  = {\widehat F^{\left( 1 \right)}}, \;\;\;\;\;\sum\limits_i {{{\bf{c}}_i}\widehat F_i^{\left( 1 \right)}}  = 0.
\label{weq_39}
\end{equation}
Summing Eqs. (\ref{eq_35}) and (\ref{eq_36}) over $i$, the corresponding equations in the $t1$ and $t2$ scales can be easily derived as 
\begin{flalign}
	&{\varepsilon ^1}:\;\;\;\;\;\rho {c_p}{\partial _{t1}}T = {\widehat F^{\left( 1 \right)}},&
	\label{eq_38}
\end{flalign} 
\begin{flalign}
	&{\varepsilon ^2}:\;\;\;\;\;\rho {c_p}{\partial _{t2}}T + \frac{{\Delta t}}{2}{\nabla _1}{\nabla _1}:c_s^2T{\bf{I}} + {\nabla _1} \cdot \sum\limits_i {{{\bf{c}}_i}h_i^{\left( 1 \right)}}  = 0.&
	\label{eq_39}
\end{flalign} 
With the aid of Eq. (\ref{weq_39}), it follows from Eq. (\ref{eq_35}) that 
\begin{equation}
	\sum\limits_i {{{\bf{c}}_i}h_i^{\left( 1 \right)}}  =  - {\tau _h}c_s^2\Delta t{\nabla _1}T.
		\label{eq_40}
\end{equation}
Using this equation, Eq. (\ref{eq_39}) further simplifies to 
\begin{equation}
\rho {c_p}{\partial _{t2}}T = {\nabla _1} \cdot \left( {{\tau _h} - \frac{1}{2}} \right)c_s^2\Delta t{\nabla _1}T.
\label{eq_41}
\end{equation}	
Combining Eq. (\ref{eq_39}) with Eq. (\ref{eq_41}) yields
\begin{eqnarray}
\rho {c_p}{\partial _t}T = \nabla  \cdot \left( {{\tau _h} - \frac{1}{2}} \right)c_s^2\Delta t\nabla T - \rho {c_p}{\bf{u}} \cdot \nabla T,
\end{eqnarray} 
with the conductivity defined as $\lambda  = \left( {{\tau _h} - 0.5} \right)c_s^2\Delta t$. Note that although there is a gradient term of $\nabla T$ appeared in the forcing term $\widehat F$, it can be locally evaluated by using Eq. (\ref{eq_40}) as 
$\nabla T = {{\sum\limits_i {\left( {{h_i} - h_i^{eq}} \right)} } \mathord{\left/{\vphantom {{\sum\limits_i {\left( {{h_i} - h_i^{eq}} \right)} } {\left( {{\tau _h}c_s^2\Delta t} \right)}}} \right.\kern-\nulldelimiterspace} {\left( {{\tau _h}c_s^2\Delta t} \right)}}$, which has been proved to be second-order accuracy in space, i.e., $\nabla T \propto \Delta {x^2}$ \cite{chai2013pre}. Form the above analysis, it can be found that the present model is able to simulate the thermocapollary flows with different heat capacities, and the main idea is to introduce a correction term of $\left( {\rho {c_p} - 1} \right){h_i}\left( {{\bf{x}} + {{\bf{c}}_i}\Delta t,t} \right)$ into the evolution equation, making the collision process is different to the classical LB method. Particularly, since the present model could avoid the calculation of the space derivatives of the heat capacitance or the order parameter, it is much simpler than some previously reported LB models. 

\section{Numerical tests}
Numerical tests are carried out in this section to show the capacity of present model in simulating thermocapillary flows. In the simulation, the standard two-dimensional nine-velocity (D2Q9) lattice \cite{kruge2017the} is adopted, which is defined as 
\begin{equation}
{{\bf{c}}_i} = \left\{ \begin{array}{l}
		c{\left( {0,0} \right)^{\rm T}},\;\;\;\;\;\;\;\;\;\;\;\;\;\;\;\;\;\;\;\;\;\;\;\;\;\;\;\;\;\;\;\;\;\;\;\;\;\;\;\;\;\;\;\;\;\;\;\;\;\;\;\;\;i = 0,\\
		c{\left( {\cos \left[ {{{\left( {i - 1} \right)\pi } \mathord{\left/
							{\vphantom {{\left( {i - 1} \right)\pi } 2}} \right.
							\kern-\nulldelimiterspace} 2}} \right],\sin \left[ {{{\left( {i - 1} \right)\pi } \mathord{\left/
							{\vphantom {{\left( {i - 1} \right)\pi } 2}} \right.
							\kern-\nulldelimiterspace} 2}} \right]} \right)^{\rm T}},\;\;\;\;\;\;\;\;\;\;\;\;i = 1,2,3,4,\\
		\sqrt 2 {\left( {\cos \left[ {{{\left( {2i - 1} \right)\pi } \mathord{\left/
							{\vphantom {{\left( {2i - 1} \right)\pi } 4}} \right.
							\kern-\nulldelimiterspace} 4}} \right],\sin \left[ {{{\left( {2i - 1} \right)\pi } \mathord{\left/
							{\vphantom {{\left( {2i - 1} \right)\pi } 4}} \right.
							\kern-\nulldelimiterspace} 4}} \right]} \right)^{\rm T}},\;\;\;\;\;i = 5,6,7,8,
	\end{array} \right.
\end{equation}
where $c = {{\Delta x} \mathord{\left/{\vphantom {{\Delta x} {\Delta t}}} \right.\kern-\nulldelimiterspace} {\Delta t}}$ is the lattice speed.  For the D2Q9 lattice, the sound speed ${c_s}$ is given as ${c_s} = {c \mathord{\left/{\vphantom {c {\sqrt 3 }}} \right.\kern-\nulldelimiterspace} {\sqrt 3 }}$, and the weight coefficient is expressed as ${\omega _0} = {4 \mathord{\left/
{\vphantom {4 9}} \right.\kern-\nulldelimiterspace} 9}$, ${\omega _{1,2,3,4}} = {4 \mathord{\left/{\vphantom {4 9}} \right.\kern-\nulldelimiterspace} 9}$ and ${\omega _{5,6,7,8}} = {1 \mathord{\left/
{\vphantom {1 {36}}} \right.\kern-\nulldelimiterspace} {36}}$. Note that there is a time derivative of ${\partial _t}\left( {\phi {\bf{u}}} \right)$ involved in the LB method for the interface-capture equation, it is computed with the first-order Euler scheme as 
\begin{equation}
{\left. {\frac{{\partial \left( {\phi {\bf{u}}} \right)}}{{\partial t}}} \right|_{\hat t}} = \frac{{{{\left. {\left( {\phi {\bf{u}}} \right)} \right|}_{\hat t}} - {{\left. {\left( {\phi {\bf{u}}} \right)} \right|}_{\hat t - \Delta t}}}}{{\Delta t}}.
\end{equation}     
To account for the unequal variables between different fluids, the physical parameter $\xi$ (such as density, dynamic viscosity, conductivity and the heat capacitance) is defined as a linear function of the order parameter
\begin{equation}
\xi (\phi )  = \frac{{\phi  - {\phi _h}}}{{{\phi _l} - {\phi _h}}}{\xi _l} + \frac{{\phi  - {\phi _l}}}{{{\phi _h} - {\phi _l}}}{\xi _h}.
\end{equation}
In a thermocapillary flow, the commonly used dimensionless numbers are given by
\begin{equation}
\mathrm{Re}=\frac{L U}{v_{h}}, \quad \mathrm{Ma}=\frac{\rho_{h} (c_p)_{h} L U}{\lambda_{h}}, \quad \mathrm{Ca}=\frac{U \mu_{h}}{\sigma_{ {ref }}},
\label{eq_46}
\end{equation} 
\begin{equation}
\rho_{r}=\frac{\rho_{l}}{\rho_{h}}, \quad \mu_{r}=\frac{\mu_{l}}{\mu_{h}}, \quad \lambda_{r}=\frac{\lambda_{l}}{\lambda_{h}}, \quad {c_{p,r}} = \frac{{{c_{p,l}}}}{{{c_{p,h}}}},
\end{equation}
where $\mathrm{Re}$, $\mathrm{Ma}$ and $\mathrm{Ca}$ are respectively the Reynolds number, Marangoni number and the capillary number with $L$ and $U$ being the characteristic length and velocity. The physical parameter with the subscript  $r$ denotes the ratio of the corresponding variable in two fluids. Note that unless otherwise stated, our numerical tests are performed by assuming the values of ${\Delta x}$ and ${\Delta t}$ are both set to be 1.0, and the interface thickness $W$ is fixed at $3.0{\Delta x}$.

\subsection{Thermocapillary flows  with two superimposed planar fluids}
\begin{figure}[H]
	\centering
	\includegraphics[width=0.5\textwidth]{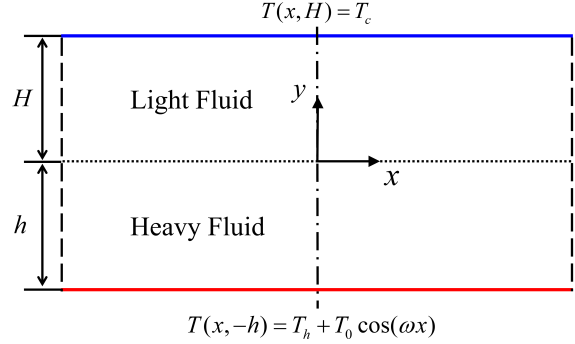}
	\caption{Schematic diagram of thermocapillary flows with two superimposed planar fluids in a heated microchannel, where $T\left( {x, - h} \right) = {T_h} + {T_0}\cos \left( {{{2\pi x} \mathord{\left/{\vphantom {{2\pi x} L}} \right.\kern-\nulldelimiterspace} L}} \right)$ and $T\left( {x,H} \right) = {T_c}$ are the temperature of the heating (lower) and cooling (upper) walls, respectively. }
	\label{fig1}
\end{figure}

We first validate the developed LB model by considering the thermocapillary flows in a heated microchannel with two superimposed planar fluids \cite{pendse2010ananalytic}. As illustrated in Fig. \ref{fig1}, the heavy fluid with the height of $h$ is located beneath the light fluid with the height of $H$. The temperature of the upper and the lower walls are $T\left( {x, - h} \right) = {T_h} + {T_0}\cos \left( {{{2\pi x} \mathord{\left/{\vphantom {{2\pi x} L}} \right.\kern-\nulldelimiterspace} L}} \right)$ and $T\left( {x,H} \right) = {T_c}$, in which  $0<T_0<T_c<T_h$. The length of the channel is set to be the characteristic length, and its valus is equal to the period of the wave number at the lower wall $L$. If the characteristic velocity of this problem is fixed at $U=\frac{|\sigma_{T}|T_0}{L} \frac{h}{\mu_{h}} $, then the governing dimensionless numbers like $Ma$, $Re$ and $Ca$ are established. Assuming the values of above three variables are sufficiently smaller than 1.0, which suggests that the convective effects induced by momentum and energy is insignificant, and the interface between two layers is able to remain flat, the corresponding analytical solutions for the upper light fluid and the lower heavy fluid the can be expressed as \cite{pendse2010ananalytic} 
\begin{flalign}
&T^{L}(x, y)=\frac{\left(T_{c}-T_{h}\right) y+\lambda_{r} T_{c} h+T_{h} H}{H+\lambda_{r} h}+T_{0} f\left(\tilde{H}, \tilde{h}, \lambda_{r}\right) \sinh (\tilde{H}-\omega y) \cos (\omega x),&
\end{flalign}
\begin{flalign}
&u_{x}^{L}(x, y)=U_{\max }\left\{\left[C_{1}^{H}+\omega\left(C_{2}^{H}+C_{3}^{H} y\right)\right] \cosh (\omega y)+\left(C_{3}^{H}+\omega C_{1}^{H} y\right) \sinh (\omega y)\right\} \sin (\omega x),&
\end{flalign}
\begin{flalign}
&u_{y}^{L}(x, y)=-\omega U_{\max }\left[C_{1}^{H} y \cosh (\omega y)+\left(C_{2}^{H}+C_{3}^{H} y\right) \sinh (\omega y)\right] \cos (\omega x),&
\end{flalign}
and
\begin{flalign}
&\begin{aligned}
	T^{H}(x, y)= \frac{\lambda_{r}\left(T_{c}-T_{h}\right) y+\lambda_{r} T_{c} h+T_{h} H}{H+\lambda_{r} h} 
	+T_{0} f\left(\tilde{H}, \tilde{h}, \lambda_{r}\right)\left[\sinh (\tilde{H}) \cosh (\omega y)-\lambda_{r} \sinh (\omega y) \cosh (\tilde{H})\right] \cos (\omega x),
\end{aligned}&
\end{flalign}
\begin{flalign}
&u_{x}^{H}(x, y)=U_{\max }\left\{\left[C_{1}^{h}+\omega\left(C_{2}^{h}+C_{3}^{h} y\right)\right] \cosh (\omega y)+\left(C_{3}^{h}+\omega C_{1}^{h} y\right) \sinh (\omega y)\right\} \sin (\omega x),&
\end{flalign}
\begin{flalign}
&u_{y}^{H}(x, y)=-\omega U_{\max }\left[C_{1}^{h} y \cosh (\omega y)+\left(C_{2}^{h}+C_{3}^{h} y\right) \sinh (\omega y)\right] \cos (\omega x).&
\end{flalign}
Note that some unknown parameters appeared in the above equations are defined as 
\begin{flalign}
&\tilde{H}=H \omega , \quad \tilde{h}=h \omega,&
\end{flalign}
\begin{flalign}
&f\left(\tilde{H}, \tilde{h}, \lambda_{r}\right)=\left[\lambda_{r} \sinh (\tilde{h}) \cosh (\tilde{H})+\sinh (\tilde{H}) \cosh (\tilde{h})\right]^{-1},&
\end{flalign}
\begin{flalign}
&\begin{aligned}
	C_{1}^{H}=\frac{\sinh ^{2}(\tilde{H})}{\sinh ^{2}(\tilde{H})-\tilde{H}^{2}} ,  C_{2}^{H}=\frac{-H \tilde{H}}{\sinh ^{2}(\tilde{H})-\tilde{H}^{2}} ,  C_{3}^{H}=\frac{2 \tilde{H}-\sinh (2 \tilde{H})}{2\left[\sinh ^{2}(\tilde{H})-\tilde{H}^{2}\right]} , \\
	 C_{1}^{h}=\frac{\sinh ^{2}(\tilde{h})}{\sinh ^{2}(\tilde{h})-\tilde{h}^{2}} , \quad C_{2}^{h}=\frac{-h \tilde{h}}{\sinh ^{2}(\tilde{h})-\tilde{h}^{2}} , \quad C_{3}^{h}=\frac{\sinh (2 \tilde{h})-2 \tilde{h}}{2\left[\sinh ^{2}(\tilde{h})-\tilde{h}^{2}\right]} ,
\end{aligned}&
\end{flalign}
\begin{flalign}
&U_{\max }=-\left(\frac{T_{0} \sigma_{T}}{\mu_{h}}\right) g\left(\tilde{H}, \tilde{h}, \lambda_{r}\right) h\left(\tilde{H}, \tilde{h}, \lambda_{r}\right),&
\end{flalign}
where 
\begin{flalign}
&g\left(\tilde{H}, \tilde{h}, \lambda_{r}\right)=\sinh (\tilde{H}) f\left(\tilde{H}, \tilde{h}, \lambda_{r}\right),&
\end{flalign}
\begin{flalign}
&h\left(\tilde{H}, \tilde{h}, \mu_{r}\right)=\frac{\left[\sinh ^{2}(\tilde{H})-\tilde{H}^{2}\right]\left[\sinh ^{2}(\tilde{h})-\tilde{h}^{2}\right]}{\mu_{r}\left[\sinh ^{2}(\tilde{h})-\tilde{h}^{2}\right][\sinh (2 \tilde{H})-2 \tilde{H}]+\left[\sinh ^{2}(\tilde{H})-\tilde{H}^{2}\right][\sinh (2 \tilde{h})-2 \tilde{h}]}.&
\end{flalign}

Numerical simulations are performed in a 2D rectangular enclosure with length and height being $160\Delta x$, $80\Delta x$, respectively. The thicknesses of the upper light fluid and the lower heavy fluid are both set to be half of the cavity height, i.e., $H=h=40\Delta x$. As for the boundary conditions, the periodic boundary conditions are applied to the horizontal direction, while the halfway bounce-back scheme is imposed to the upper and lower walls. Note that the influence of the wall wettability is not taken into account in our simulations. Additionally, all physical parameters used in this problem are the same as previous works, in which the wall temperatures are set according to $T_h=20$, $T_c=10$, $T_0=4$, $T_{ref}=10$, and fluid properties are given by  $\sigma_{ {T }}=-5\times10^{-4}$, $\sigma_{ {ref }}=2.5\times10^{-2}$, $\rho_{l}=\rho_{h}=1.0$, $(c_p)_{l}=(c_p)_{h}=1.0$, $\mu_{l}=\mu_{h}=0.2$, $\lambda_{h}=0.2$.  Further, to investigate the effect of the thermal conductivity ratio on the induced thermocapillary flows and the resulting temperature distribution, we consider two distinct values of the thermal conductivity ratio: namely, $\lambda_{r}=1.0$ and  $\lambda_{r}=0.2$, such that $\lambda_{l}=0.2$ and $\lambda_{l}=0.04$.

\begin{figure}[H]
	\centering
	\subfigure[]{\label{fig4a}
		\includegraphics[width=0.46\textwidth]{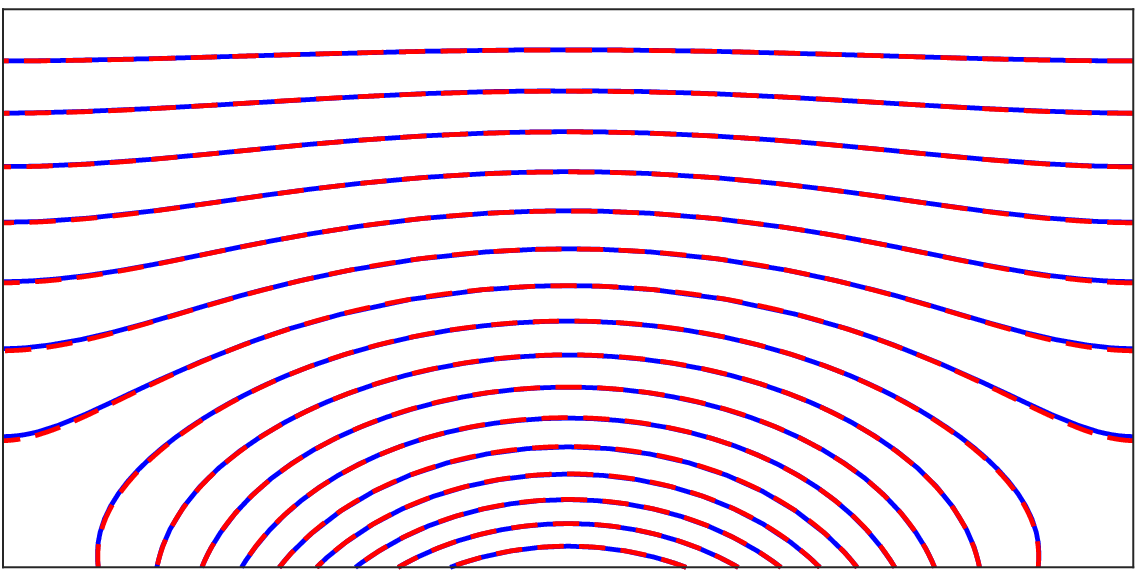}
	}
	\subfigure[]{\label{fig4b}
		\includegraphics[width=0.46\textwidth]{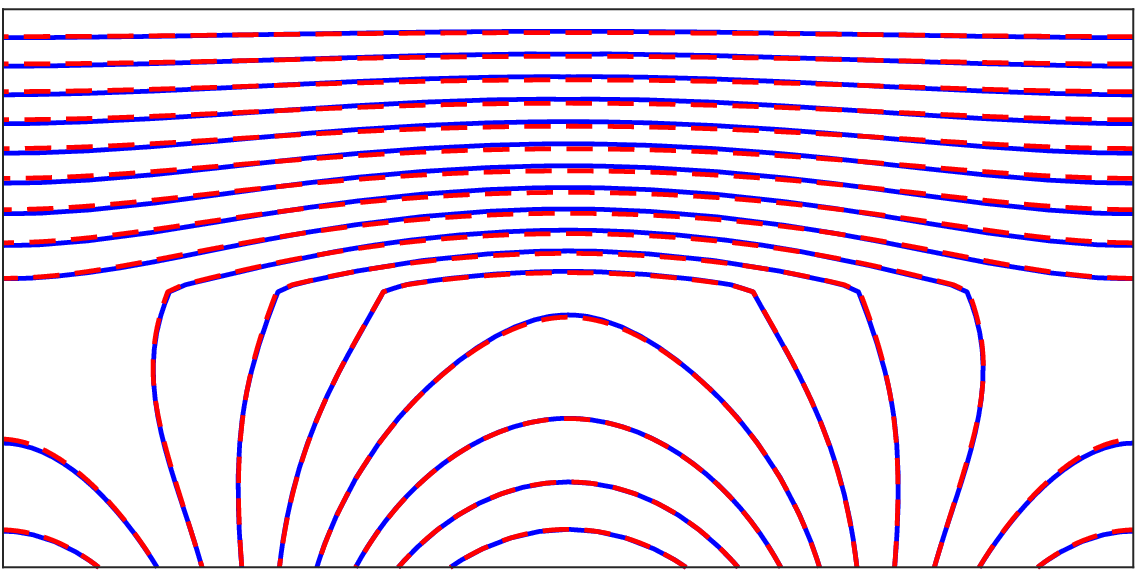}
	}
	\caption{Isothermals of thermocapillary flows with two superimposed planar fluids at (a) $\lambda_r=1.0$ and (b) $\lambda_r=0.2$, where the analytical and numerical results are denoted by the red-solid and blue-dashed lines with arrow and, respectively. }
	\label{fig2}
\end{figure}

\begin{figure}[H]
	\centering
	\subfigure[]{\label{fig4a}
		\includegraphics[width=0.46\textwidth]{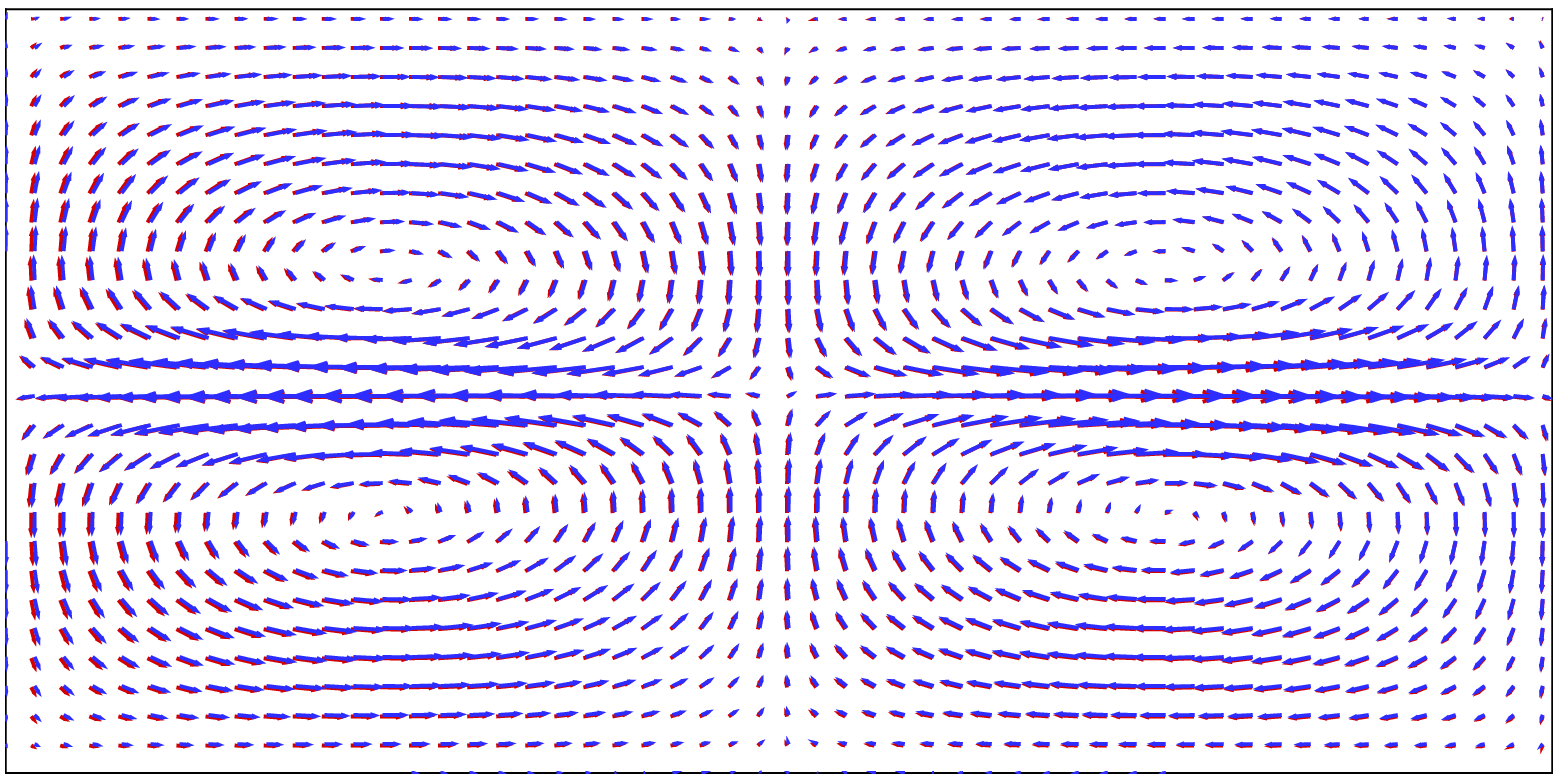}
	}
	\subfigure[]{\label{fig4b}
		\includegraphics[width=0.46\textwidth]{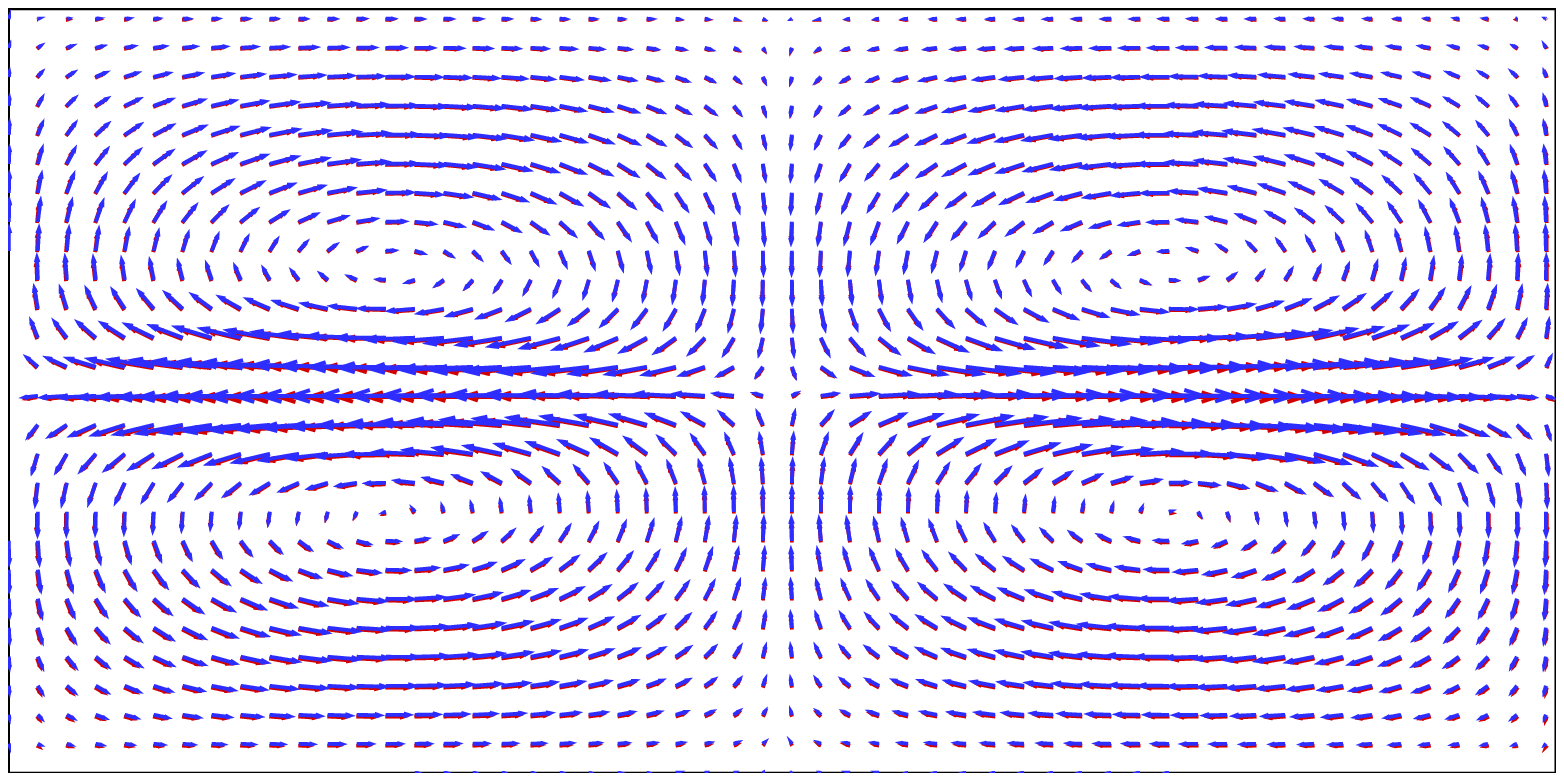}
	}
	\caption{Velocity vectors of thermocapillary flows with two superimposed planar fluids at (a) $\lambda_r=1.0$ and (b) $\lambda_r=0.2$, where the analytical and numerical results are denoted by the red and blue lines with arrow and, respectively. }
	\label{fig3}
\end{figure}

\begin{figure}[H]
	\centering
	\subfigure[]{\label{fig4a}
		\includegraphics[width=0.46\textwidth]{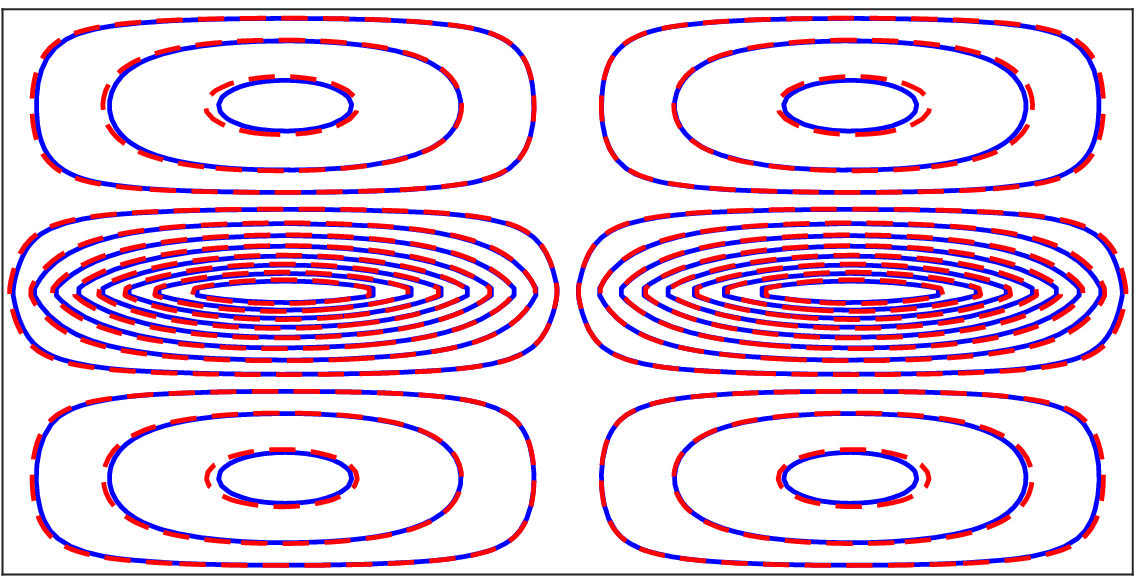}
	}
	\subfigure[]{\label{fig4b}
		\includegraphics[width=0.46\textwidth]{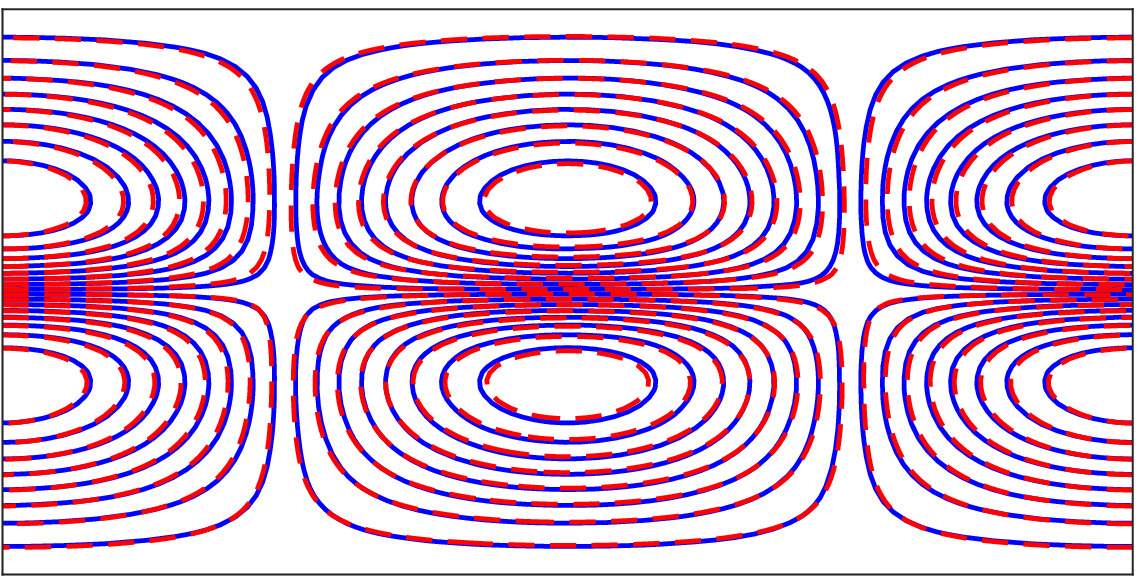}
	}
	\caption{Contours of horizontal velocity $u_x$ (a) and  vertical velocity $u_y$ (b) for thermocapillary flows with two superimposed planar fluids at $\lambda_r=0.2$, , where the analytical and numerical results are denoted by the red-solid and blue-dashed lines with arrow and, respectively. }
	\label{fig4}
\end{figure}

Fig. \ref{fig2} depicts the distributions of the isothermals for different thermal conductivity ratios obtained by the present LB approach. In order to assess the numerical performance, the analytical solutions for predicting the isothermals are also plotted in this figure. From Fig. \ref{fig2}, it can be clearly seen that our numerical solution approaches the analytical solution very well even at unequal thermal conductivity ratio (i.e., $\lambda_r=0.2$), while it is noted that the isothermals predicted by Liu et al.'s model \cite{liu2012modeling} slightly deviate from the analytical solution under the same condition. This suggests that the present LB model for thermocapillary flows is able to predict temperature with relatively small deviations. Additionally, it is interesting to observe that when there is a jump of thermal conductivity across the interface, the isotherms at the upper fluid are nearly parallel to each other, while they tend to be normal to the interface for the lower fluid. As far as the velocity field is considered, Fig. \ref{fig3} shows the comparisons of the numerical results of the present LB scheme with the analytical solutions for the velocity vectors which shows that the present numerical results are in good agreement with the analytical soultions. 

\begin{table*}[htb]
	\caption{Comparisons of the relative errors of the temperature (T), the horizontal velocity ($u_x$), and the vertical velocity ($u_y$) at different d thermal conductivity of fluids. Here, the relative error is defined as ${E_\delta } = \sqrt {{{\sum\limits_{\bf{x}} {{{\left| {{\xi _n} - {\xi _a}} \right|}^2}} } \mathord{\left/{\vphantom {{\sum\limits_{\bf{x}} {{{\left| {{\xi _n} - {\xi _a}} \right|}^2}} } {\sum\limits_{\bf{x}} {{{\left| {{\xi _a}} \right|}^2}} }}} \right.
					\kern-\nulldelimiterspace} {\sum\limits_{\bf{x}} {{{\left| {{\xi _a}} \right|}^2}} }}} $, where ${{\xi _a}}$  and ${{\xi _n}}$ represent the analytical solution and the numerical result of the parameter $\xi$, respectively. .}
	\label{table1}
	\centering
	\setlength{\tabcolsep}{3mm}{
		\begin{tabular}{ccccccc}
			\hline
			\hline	
			&	$\lambda_r$ 	& Relative Error  & Present                 & Liu et al\cite{liupre2013phase}            & Mitchell et al. \cite{mitchell2021compu}          & Yue et al. \cite{yue2022improved}  \\
			\hline
			&0.2	            &  $u_x$          &  $6.08\times 10^ {-2}$  &  -                     &  -                       &  $4.91\times 10^ {-2}$ \\
			&	                &  $u_y$          &  $7.00\times 10^ {-2}$  &  $8.41\times 10^ {-2}$ &  $1.42\times 10^ {-1}$   &  $5.90\times 10^ {-2}$ \\
			&	                &  $T$            &  $1.40\times 10^ {-3}$  &  $5.23\times 10^ {-3}$ &  $7.65\times 10^ {-3}$   &  $6.73\times 10^ {-3}$ \\
			&1.0                &  $u_x$          &  $3.66\times 10^ {-2}$  &  -                     &  -                       & $4.35\times 10^ {-2}$\\
			&	                &  $u_y$          &  $4.47\times 10^ {-2}$  &  $5.71\times 10^ {-2}$ &  $1.18\times 10^ {-1}$   & $5.80\times 10^ {-2}$\\
			&	                &  $T$            &  $4.37\times 10^ {-4}$  & $2.25\times 10^ {-4}$  &  $8.27\times 10^ {-4}$   & $3.06\times 10^ {-4}$\\

			\hline
			\hline
	\end{tabular}}
\end{table*}

In previous work by Yue et al. \cite{yue2022improved} , the authors showed that the vertical velocities of two fluids near the interface obtained by their approach at $\lambda_r=1.0$ are asymmetry and fluctuant, which is an unfavorable behavior. This unphysical phenomenon implies that the numerical performance of their model around the multiphase interface may be not very well in certain cases. In contrast, from Fig. \ref{fig4} we can clearly observe that even at $\lambda_r=0.2$ (a more general circumstance),  both the horizontal and vertical velocities obtained by the present scheme are smooth in the whole cavity, and they are also consistent with the theoretical prediction, implying that the numerical stability of our scheme is better than that of Yue et al. \cite{yue2022improved}. Finally, to quantify the results, Table \ref{table1} gives the relative numerical errors between the numerical results and the analytical solutions for the horizontal velocity $u_x$, the vertical velocity $u_y$ and the temperature $T$. It is seen that the errors predicted by Mitchell et al.'s model\cite{mitchell2021compu} are the biggest, and the errors of our model and Yue et al. are comparable, and they are generally smaller than those of Liu et al. \cite{liupre2013phase}.

\subsection{Thermocapillary migration of deformable droplets}
Thermocapillary migration of deformable droplets is a classical problem for investigating the Marangoni effect, which also has been widely used to assess the proposed multiphase LB methods for simulating thermocapillary flows. To our knowledge, however, most of previous studies only focus on the thermocapillary migration of deformable droplets with equal specific-heat capacities \cite{liu2012modeling,liujcp2015modelling,liu2017alattice,zheng2016continuous}, and the models cannot be directly applied to  thermocapillary flows with heat capacitances contrasts. In this subsection, we will adopt the present LB method to simulate the thermocapillary migration of deformable droplets with different thermophysical parameters. Before addressing the initial setup of this physical problem, we briefly introduce the theoretical solution for this problem.  In 1959, Young et al. \cite{young1959the} first analyzed the thermocapillary migration in the case of infinitesimal  Reynolds and Marangoni numbers, in which the convection transport of momentum and energy is so weak that they can be neglected in contrast to the molecular transport of these quantities, and the governing equations can then be linearized. In this setting, the steady migration velocity of spherical droplet within an unbounded fluid medium for two fluids of equal thermal conductivity can be expressed as \cite{young1959the}    
\begin{equation}
	U_{YGB}=\frac{2U}{(2+3\mu_{r})(2+\lambda_{r})},
	\label{eq_60}
\end{equation}
where $U = {{ - {\sigma _T}\nabla {T_\infty }R} \mathord{\left/{\vphantom {{ - {\sigma _T}\nabla {T_\infty }R} {{\mu _h}}}} \right.
\kern-\nulldelimiterspace} {{\mu _h}}}$ is the characteristic  thermocapillary velocity with $R$ and ${\nabla {T_\infty }}$ being the droplet radius and the temperature gradient imposed in the system, respectively. 

\begin{figure}[h]
	\centering
	\includegraphics[width=0.3\textwidth]{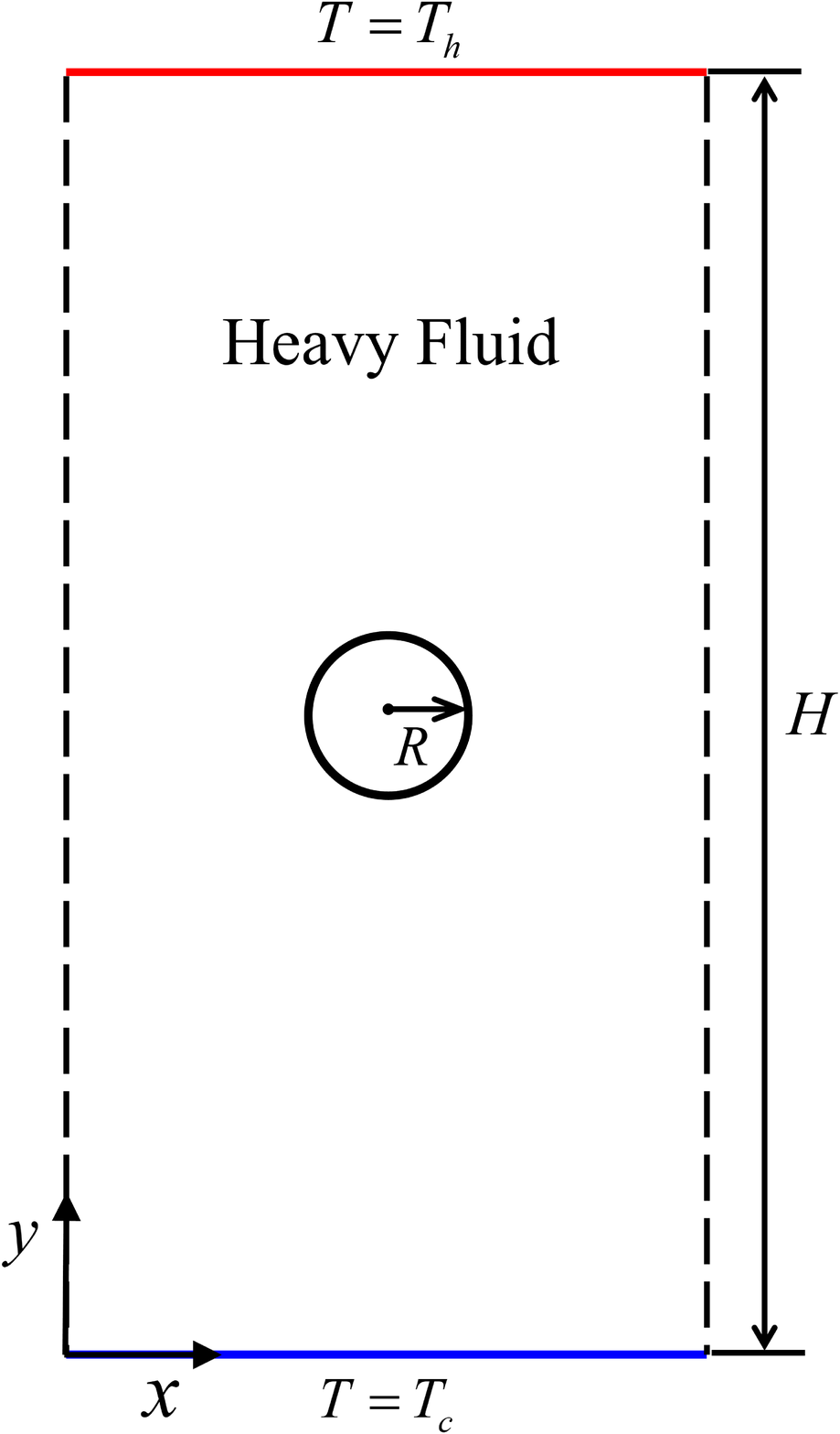}
	\caption{The physical sketch of the thermocapillary migration of a deformable droplet.}\label{fig4}
\end{figure}

\begin{figure}[H]
	\centering
	\includegraphics[width=0.5\textwidth]{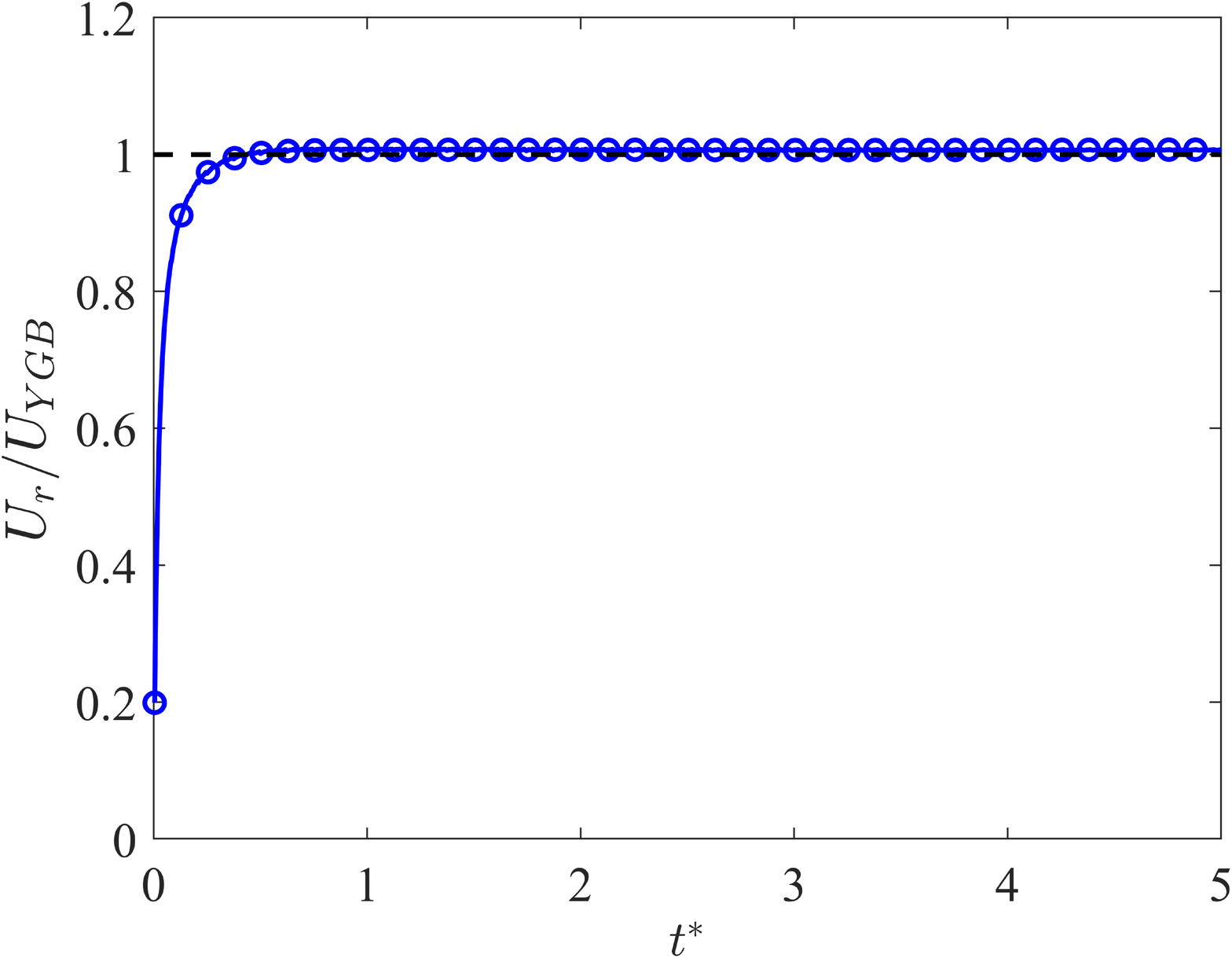}
	\caption{Time evolution of normalized migration velocity of a droplet at $Ma=Re=0.1$, $Ca=0.08$, where the analytical solution and the numerical results are denoted by the dashed line and the solid line.}\label{fig5}
\end{figure}
Fig. \ref{fig4} shows the physical sketch of the current problem. In the LB simulation, the grid resolution is set to be  $160 \Delta x \times 320 \Delta x$ with the periodic boundary conditions in the horizontal direction \cite{kruge2017the,guobook2013lattice}. The bottom and top planes are two solid walls imposed by the non-slip bounce back boundary condition \cite{kruge2017the,guobook2013lattice}. Initially, a droplet of radius $R=20 \Delta x$ is placed at the center of the cavity, and a linear temperature field is posed in the vertical direction with $T_h=32$ on the top wall and $T_c=16$ ($T_{ref}$) on the bottom wall, resulting in $|\nabla {T}_\infty|=0.1$. We first test the accuracy of the present LB scheme by setting $\sigma_{ {ref }}=2.5\times 10^{-3}$, $\sigma_{ {T }}=-10^{-4}$, $\rho_{l}=\rho_{h}=1$, $c_{p,l}=c_{p,h}=1.0$, $\mu_{l}=\mu_{h}=0.2$, $\lambda_{l}=\lambda_{h}=0.2$. Based on Eq. (\ref{eq_46}) and Eq. (\ref{eq_60}), the theoretical steady migration velocity $U_{YGB}$ is $1.333\times10^{-4}$ , the Reynolds $\rm{Re}$ and Marangoni numbers $\rm{Ma}$ are both equal to 0.1 and the Capillary number $Ca$ is $0.08$. In addition, we use the following equation to  account for the droplet velocity \cite{liu2012modeling},  
\begin{equation}
	U_{r}(t)=\frac{\int_{V} \phi u_{y} d V}{\int_{V} \phi d V}=\frac{\sum_{\mathbf{x}} \phi(\mathbf{x}, t) u_{y}(\mathbf{x}, t)}{\sum_{\mathbf{x}} \phi(\mathbf{x}, t)}, \quad \phi<0.
\end{equation}
Fig. \ref{fig5} present the time evolution of the migration velocity numerically predicted by the present method, in which the time has been normalized by ${R \mathord{\left/{\vphantom {R U}} \right.\kern-\nulldelimiterspace} U}$. Based on this figure,  it is clear that our scheme can provide satisfactory predictions for this test case.

\begin{figure}[H]
	\centering
	\includegraphics[width=0.5\textwidth]{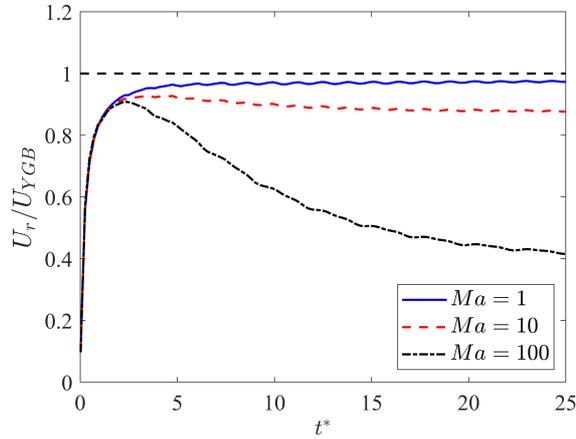}
	\caption{Time evolutions of the droplet migration velocity at different Marangoni number at $Re=1$.}
	\label{figW7}
\end{figure}

The above simulations are conducted in the limit of zero small Marangoni number, however, this is not generally the case. In many natural physical processes particularly in space material processing, studies on thermocapillary flows with finite Marangobi numbers attract much more interest from researchers in the corresponding fields \cite{balasubramaniam2000the,xie2005experiment}. To this end, we also study the same problem for four different Marangoni numbers, i.e., $Ma=1, 10$ and 100,  and the Reynolds number is fixed at $Re=1.0$. Note that different values of Marangoni numbers are achieved by modifying the $c_p$ whilst keeping $\lambda_{l}=\lambda_{h}=0.1$. In addition, apart from $\mu_{l}=\mu_{h}=0.1$, $\sigma_{ {ref }}=5\times 10^{-3}$, $\sigma_{ {T }}=-2.5\times10^{-4}$, all the other physical parameters are the same as those used in the previous test. Fig. \ref{figW7} shows the time evolutions of the normalized migration velocity for different values of $\rm{Ma}$. It is observed that due to the effect of the initial condition, i.e., ${\left. {\bf{u}} \right|_{t = 0}} = 0$ and ${\left. T \right|_{t = 0}} = y\nabla {T_\infty }$,  the droplet migration velocity at the early stage for all cases are increased with the same speed. After that, the migration velocity of $Ma=1$ is the first to reach a steady value, while for a relatively large value of $\rm{Ma}$ ($\rm{Ma} \ge 10$), the migration velocity is found to decrease before reaching the steady state.  Also, it is shown that the terminal migration velocity for different Marangoni numbers decreases with $Ma$, and this observation fits well with previous theoretical and numerical findings for the case of nondeformable drops \cite{shankar1988the,yin2008thermo}. To have a better understanding on this point, we also present the distributions of the streamlines and isotherms in Fig. \ref{figW7}. It is clear that the isotherms are nearly parallel to each other when $Ma$ takes 1.0, indicating that the heat transfer in this case is mainly transport by diffusion. As $Ma$ increases, however, the isotherms around the droplet bend upward suggesting that the convective transport of energy in the system is enhanced. Further, it is noticed that the intensity of the vortex decreases with increasing $Ma$ (see  Fig. \ref{figW7}). Actually, due to the temperature gradient at the droplet surface is decreased with $Ma$, the droplet driving force in the vertical direction is expected to be reduced, which is just the reason why the terminal migration velocity (also the vortex intensity) decreases with $Ma$. Finally, we would like to investigate the numerical performance of the present LB scheme at high density ratio. To this end, we simulate this problem with four different density ratios, i.e., ${\rho _r}=10, 100, 500$ and 1000, and the fluid properties are set to be ${\sigma _{ref}} = 2.5 \times {10^{ - 4}}$, ${\sigma _T} =  - 1.0 \times {10^{ - 5}}$, ${\mu _l} = {\mu _h} = 0.8$, ${\left( {\rho {c_p}} \right)_h} = 5$ and ${\left( {\rho {c_p}} \right)_l} = 1.0$, leading to ${\mathop{\rm Re}\nolimits}  = 0.00625$, ${\rm{Ma}} = 0.0125$, ${\rm{Ca}} = 0.08$, ${\Pr _h} = 20$ and ${\Pr _l} = 8$ (here, $\rm{Pr}$ is the Prandtl number defined as  $\Pr  = {{{c_p}\mu } \mathord{\left/{\vphantom {{{c_p}\mu } \lambda }} \right.\kern-\nulldelimiterspace} \lambda }$). All other unmentioned parameters are the same as those used the above tests cases. Table \ref{table2} gives the terminal migration velocities obtained by the present LB scheme at various density-ratios, where the analytical solutions $U_{YGB}$ are also included. It is clear that our numerical results agree well with the analytical solutions, indicating that our scheme is able to simulate thermocapillary flows with larger density contrast.       

\begin{figure}[H]
	\centering
	\begin{minipage}[c]{0.15\textwidth}
		\centering
		\caption*{(a) $Ma=1$}
	\end{minipage}
	\begin{minipage}[c]{0.42\textwidth}
		\includegraphics[width=\textwidth]{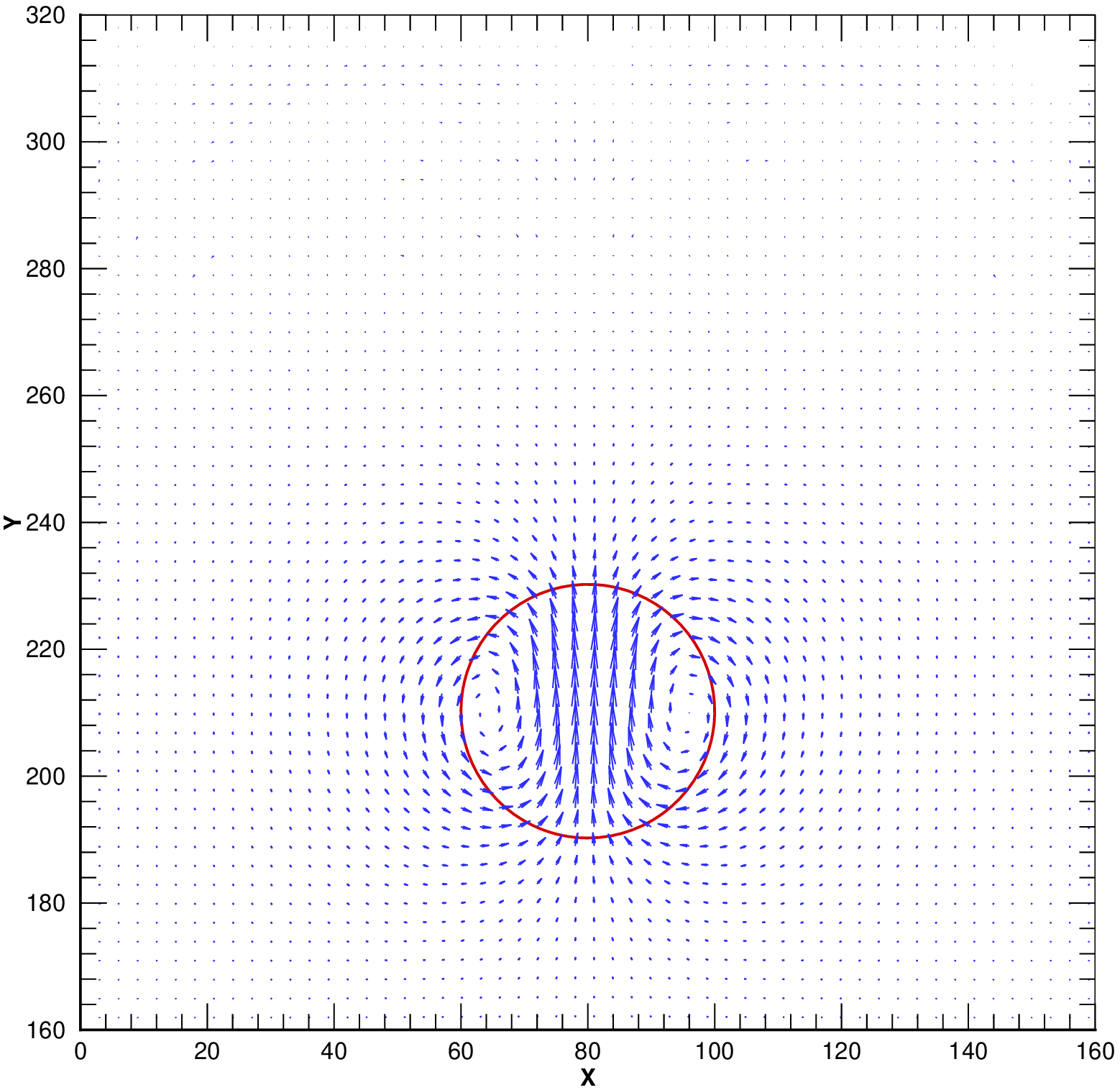}
	\end{minipage}
\begin{minipage}[c]{0.42\textwidth}
	\includegraphics[width=\textwidth]{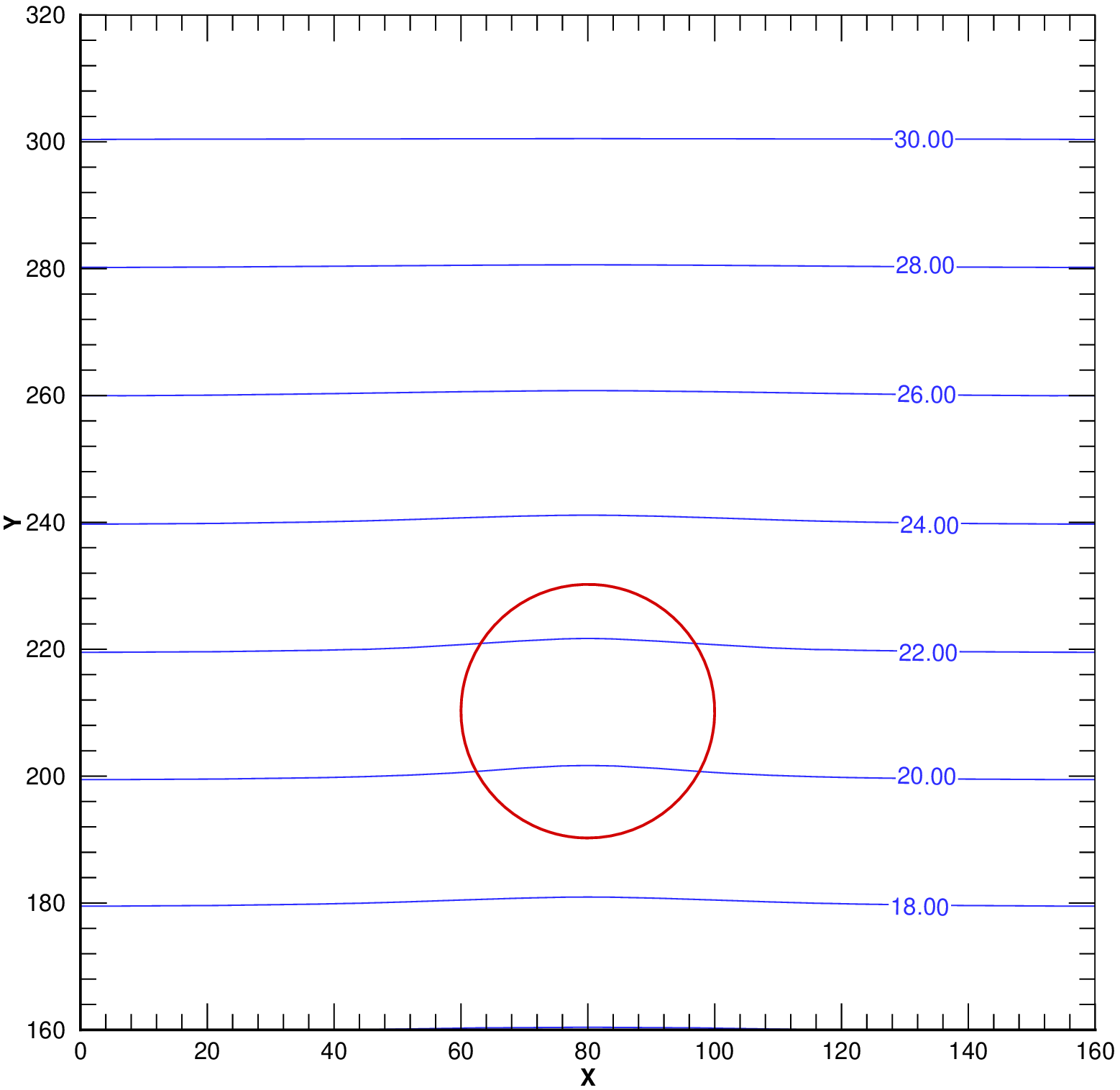}
\end{minipage}
	\begin{minipage}[c]{0.15\textwidth}
		\centering
		\caption*{(b) $Ma=10$}
	\end{minipage}
\begin{minipage}[c]{0.42\textwidth}
	\includegraphics[width=\textwidth]{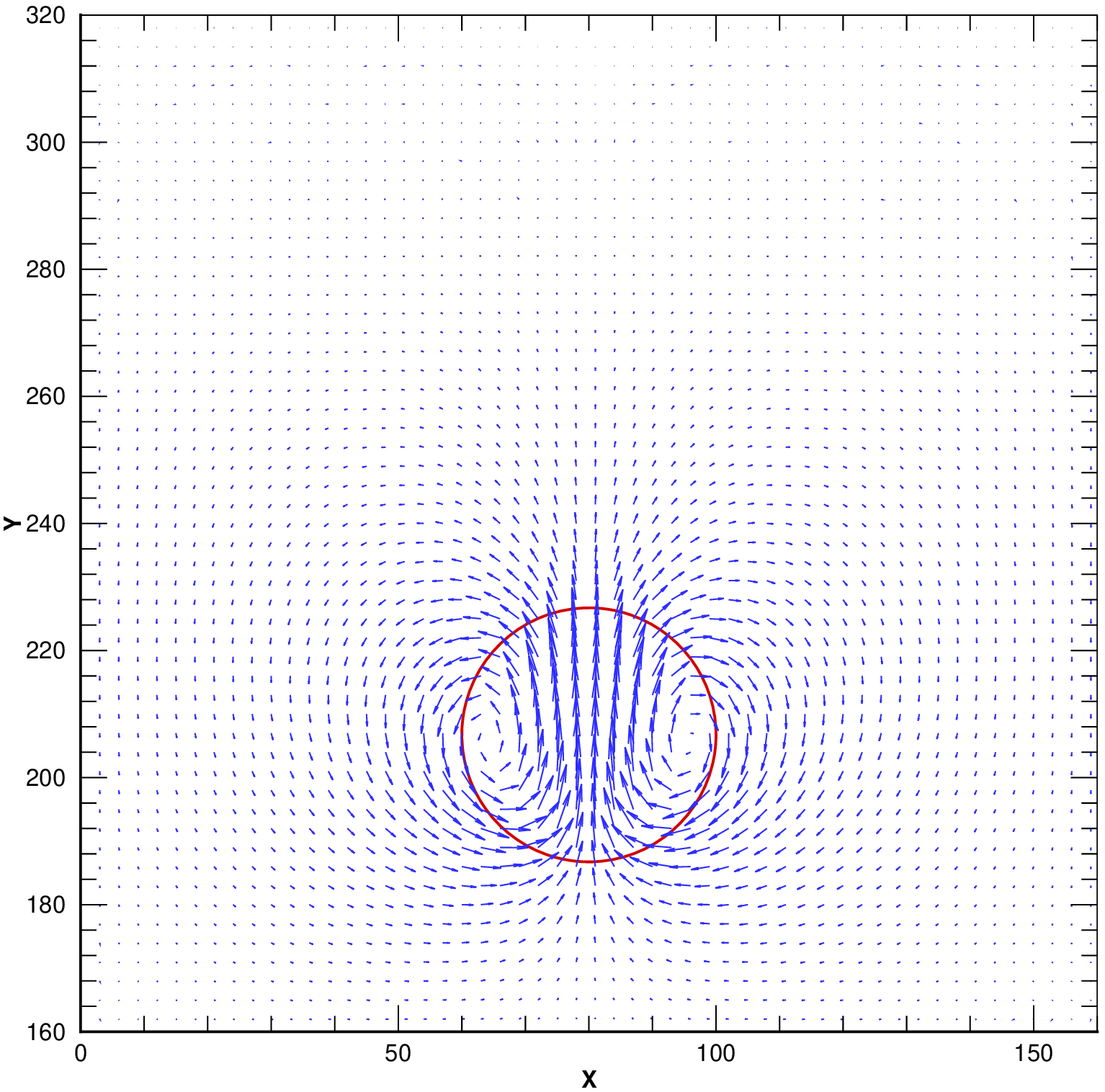}
\end{minipage}
\begin{minipage}[c]{0.42\textwidth}
	\includegraphics[width=\textwidth]{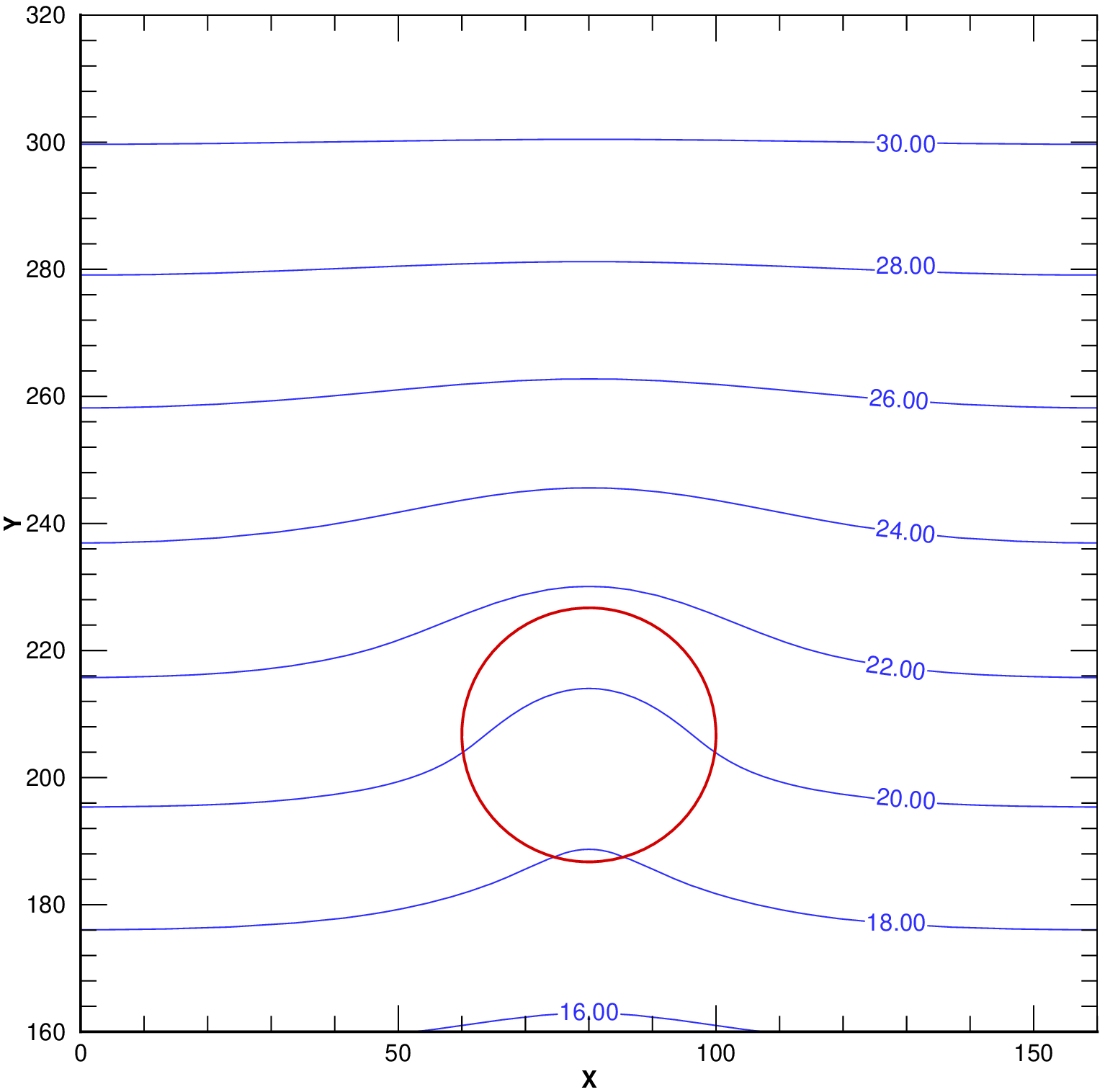}
\end{minipage}
	\begin{minipage}[c]{0.15\textwidth}
		\centering
		\caption*{(c) $Ma=100$}
	\end{minipage}
	\begin{minipage}[c]{0.42\textwidth}
		\includegraphics[width=\textwidth]{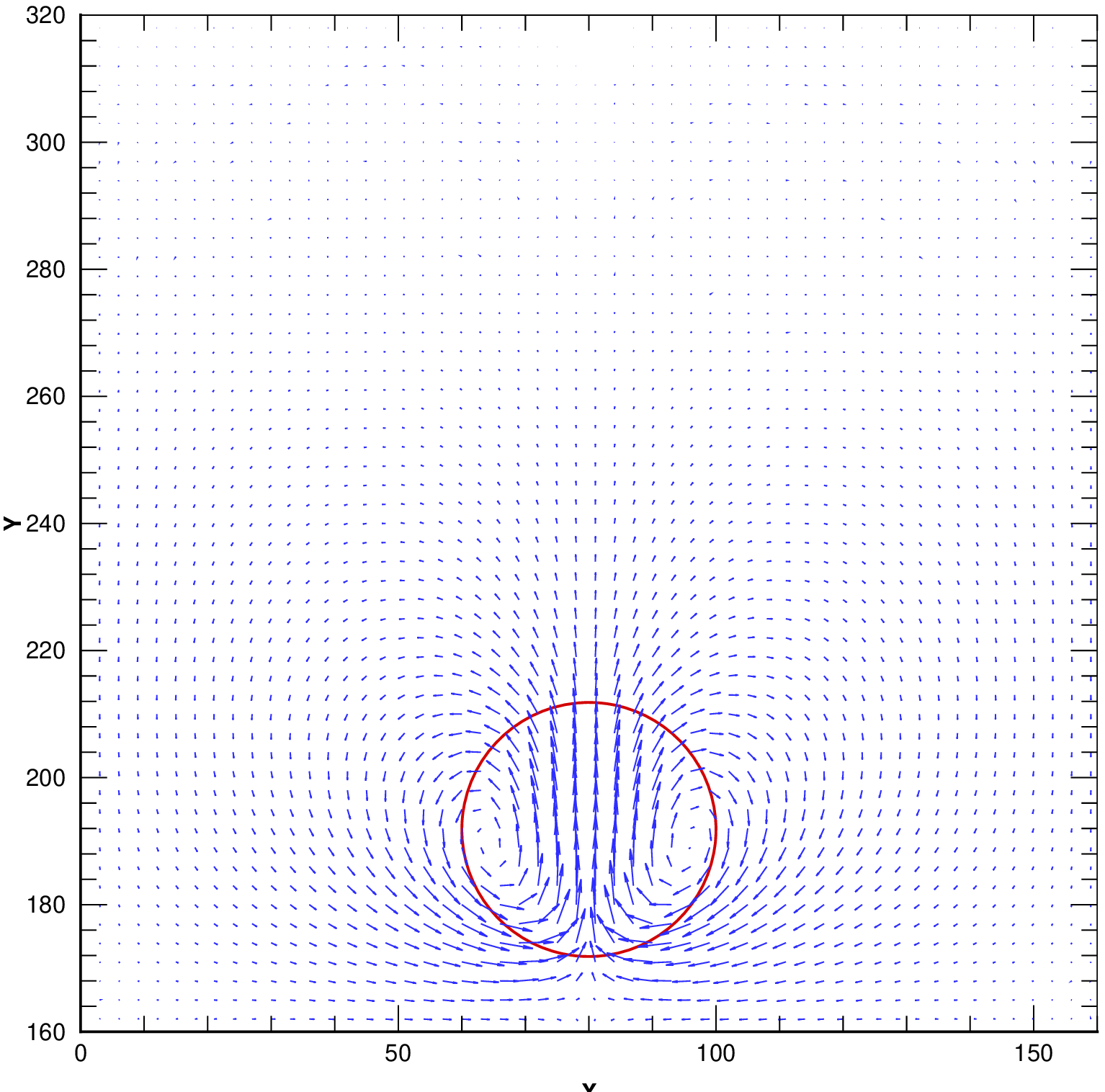}
	\end{minipage}
	\begin{minipage}[c]{0.42\textwidth}
		\includegraphics[width=\textwidth]{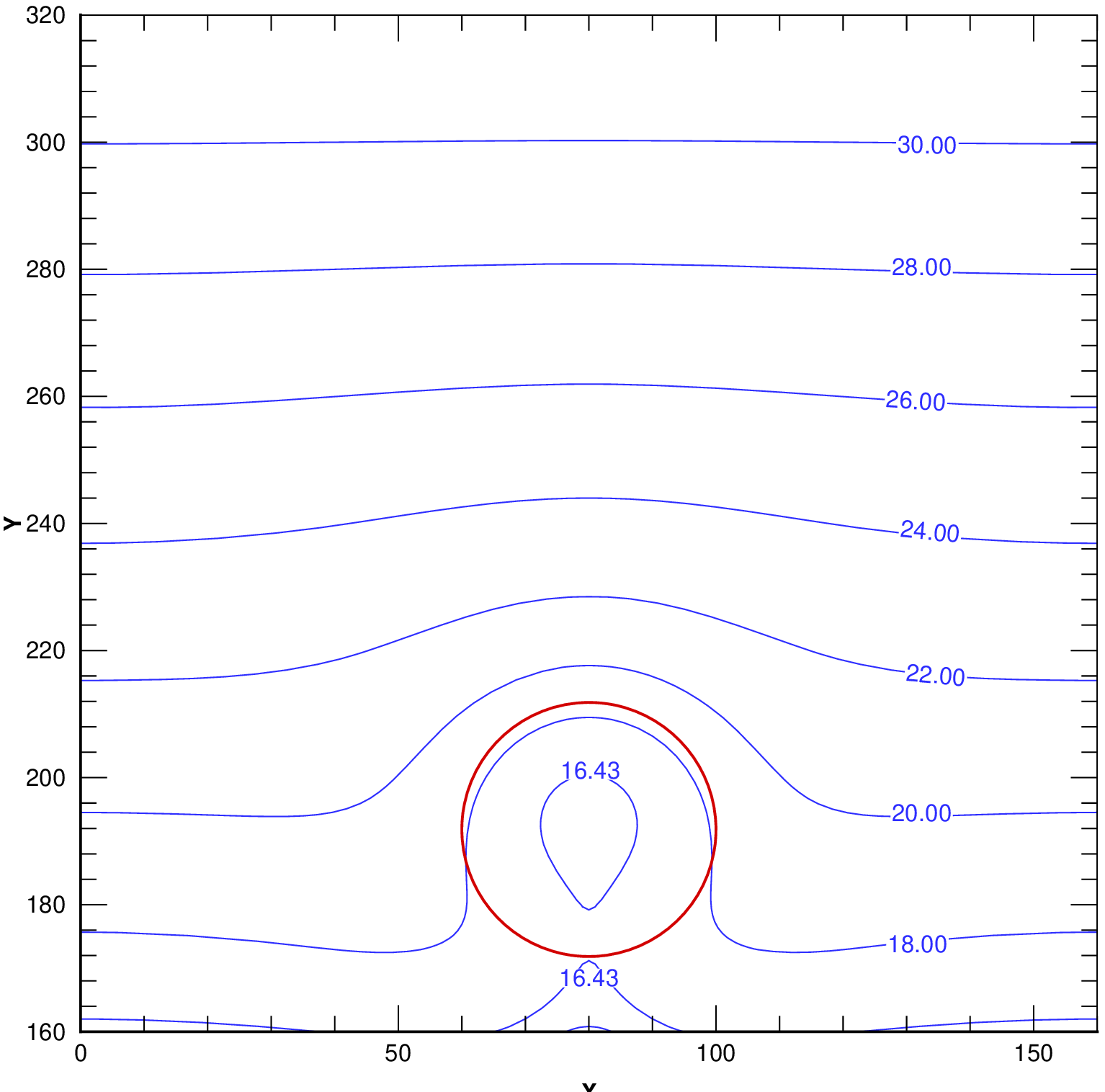}
	\end{minipage}
	\caption{Velocity vectors (left) and isotherms (right) around the migrating droplet at (a) $Ma=1$, (b) $Ma=10$, (c) $Ma=100$.}
	\label{figW7}
\end{figure}

\begin{table*}[htb]
	\caption{Comparisons of the terminal migration velocity between theory solution $U_{YGB}$ and the numerical results $U_r$ for four different density ratios.}
	\label{table2}
	\centering
	\setlength{\tabcolsep}{3mm}{
		\begin{tabular}{lcccccc}
			\hline
			\hline	
			&	$\rho_r$ 	    &  $U_{YGB}$                         &   & $U_{r}$                     &    & Relative error \\
			\hline                 
			&1:10	            &  $8.6956\times 10^ {-6}$          &   &  $8.3296\times 10^ {-6}$   &    &  $4.39\% $ \\
			&1:100	            &  $9.8522\times 10^ {-6}$          &   &  $9.5978\times 10^ {-6}$   &    &  $2.58\% $ \\
			&1:500	            &  $9.9700\times 10^ {-6}$          &   &  $9.7345\times 10^ {-6}$   &    &  $2.39\% $ \\
			&1:1000             &  $9.9850\times 10^ {-6}$          &   &  $9.7528\times 10^ {-6}$   &    &  $2.35\% $\\
			\hline
			\hline
	\end{tabular}}
\end{table*}

\subsection{Thermocapillary flow of two recalcitrant bubbles}
The above numerical tests assume a linear relationship between the surface tension and temperature, however, it has been reported that the surface tension may be a parabolic dependence of the temperature in some special cases such that there is a maximum or minimum surface tension when varying temperature \cite{shanahan2014reca}. In 2015, Shanahan and Sefiane \cite{shanahan2014reca} experimentally  studied the thermocapillary motion of bubbles in a channel by considering the surface tension to be a parabolic function of the temperature. To their surprise, the droplet near the channel outlet are inclined to move against the liquid flow until it reaches equilibrium at a specific point. Based on this work, thereafter Majidi et al. \cite{majidi2020single} numerically investigated the migration of a single recalcitrant bubble, in which the effects of the density ratio, viscosity ratios, inlet velocity are studied in detail. Note that although the Allen-Cahn equation in Majidi et al.'s work \cite{majidi2020single}  is solved by the LB method, the temperature field is still predicted by the finite difference scheme. In this subsection, to further verifies the applicability of the current method, we intend to study the similar problem, and the difference between our and previous simulations lies in that there are two recalcitrant bubbles in the channel. 
\begin{figure}[H]
	\centering
	\includegraphics[width=0.5\textwidth]{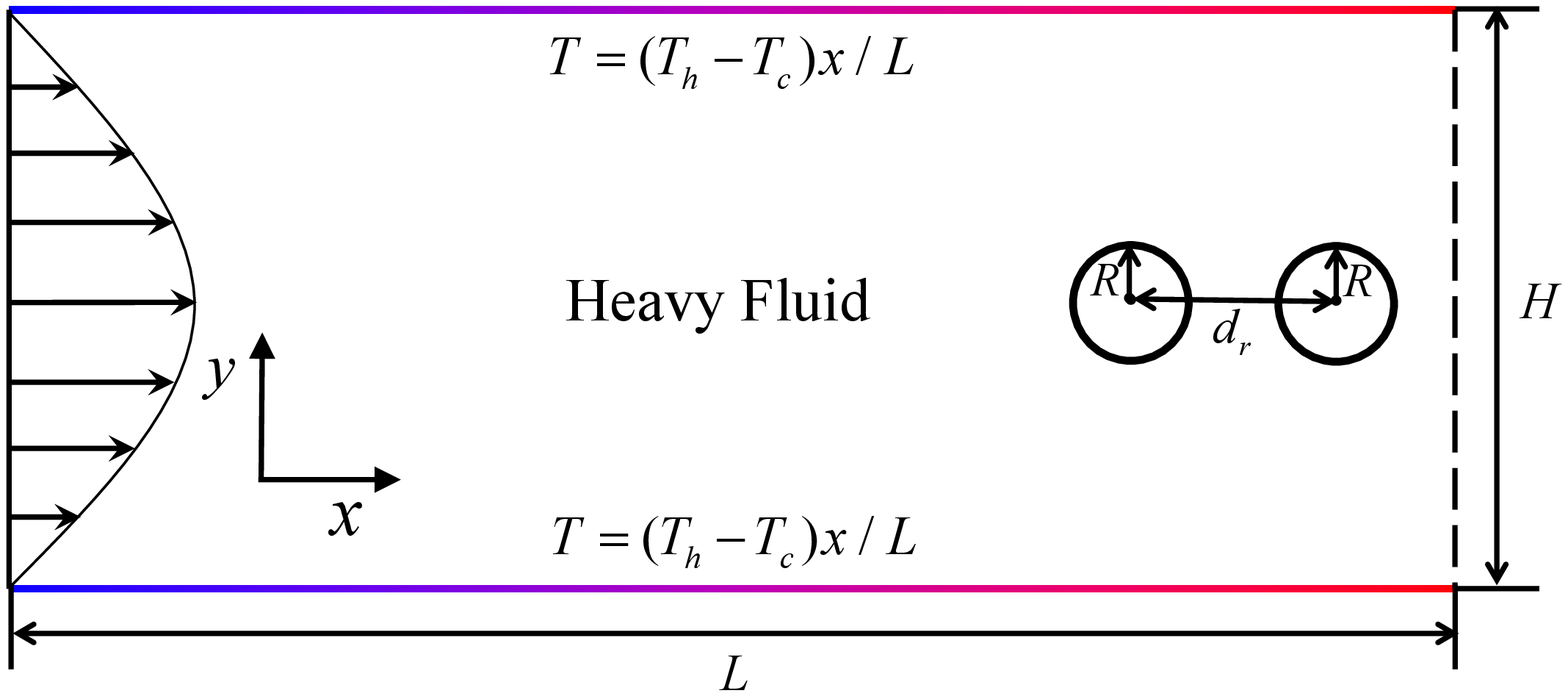}
	\caption{The schematic diagram of two recalcitrant bubbles in a micro-channel .}
	\label{figw9}
\end{figure}

The schematic of the problem is depicted in Fig. \ref{figw9}, and our simulations are performed in a rectangular microchannel with $L=500 \Delta x$ and $H= 200 \Delta x$.  Two bubbles with the radius of $R=20\Delta x$ and the distance of $d_r$ are initially located in the downstream of the channel, and the center of the bubble near the outlet is just placed at $(460 \Delta x, 100 \Delta x)$. In the simulation, a fully developed velocity profile is applied to the inlet boundary, which can be described as 
\begin{equation}
	{u_y}\left( y \right) = 6{u_{ave}}\frac{{y\left( {H - y} \right)}}{{{H^2}}},
\end{equation}
where $u_{ave}$ is the average inlet velocity. The no-slip bounce-back boundary condition is imposed at the top and the bottom boundaries, and the temperature on these two solid walls are specified through ${{\left( {{T_h} - {T_c}} \right)x} \mathord{\left/{\vphantom {{\left( {{T_h} - {T_c}} \right)x} L}} \right.\kern-\nulldelimiterspace} L}$. The inlet is kept at a constant cold temperature of $T_c$, and the adiabatic boundary condition is applied to the outlet boundary, where the flow information is obtained by the stress-free boundary condition. Note that to prevent mass leakage in the horizontal direction, the classical boundary condition proposed by Zou and He \cite{zou1997on}
is modified here, and the details can be found in Ref. \cite{liu2012modeling}. Additionally, following previous works, the surface tension used in the simulation is assume to be a parabolic function of the temperature, which can be defined as 
\begin{equation}
\sigma \left( T \right) = \frac{1}{2}{\sigma _{TT}}{\left( {T - {T_{ref}}} \right)^2} + {\sigma _{ref}},
\label{eq_63}
\end{equation}
where ${\sigma _{TT}}$ is the second derivative of the surface tension, i.e., ${\sigma _{TT}} = {{{\partial ^2}\sigma } \mathord{\left/{\vphantom {{{\partial ^2}\sigma } {\partial {T^2}}}} \right.\kern-\nulldelimiterspace} {\partial {T^2}}}$, and it is clear that the minimum value of the surface tension equals ${\sigma _{ref}}$ when ${T = {T_{ref}}}$.
The parameters used in the simulation are chosen as ${\nu _h} = 0.04$, ${\nu _l} = 0.2$, ${d _r} = 4.0R$, ${u_{ave}} = 2.5 \times {10^{ - 4}}$, ${\rho _h} = 1.0$, ${\rho _l} = 0.1$, ${c_{p,h}} = {c_{p,l}} = 20.0$, ${\lambda _h} = 7.0$ and ${\lambda _l} = 0.7$.  With the driving of the thermocapillary force, the droplets are inclined to move towards the hotter areas of the flows, and induce the interaction of the two bubbles. To illustrate the migration characteristics of the recalcitrant bubbles in the microchannel, the migration velocity and the local position of each bubble are determined through \cite{mitchell2021compu} 
\begin{equation}
	u_{b, \alpha}=\frac{\iint_{\phi_\alpha<0} u_{x} \phi d x d y}{\iint_{\phi_\alpha<0} \phi d x d y},\;\;\;\; x_{b, \alpha}=\frac{\iint_{\phi_\alpha<0} x \phi d x d y}{\iint_{\phi_\alpha<0} \phi d x d y}
\end{equation}
where the subscript $\alpha$ refers to the bubble of interest.  In the following simulation, we mainly concentrate on the interfacial dynamics and the variation of the migration behavior versus time under different  ${\sigma _{TT}}$.   

\begin{figure}[H]
	\centering

	\begin{minipage}[c]{0.49\textwidth}
		\includegraphics[width=\textwidth]{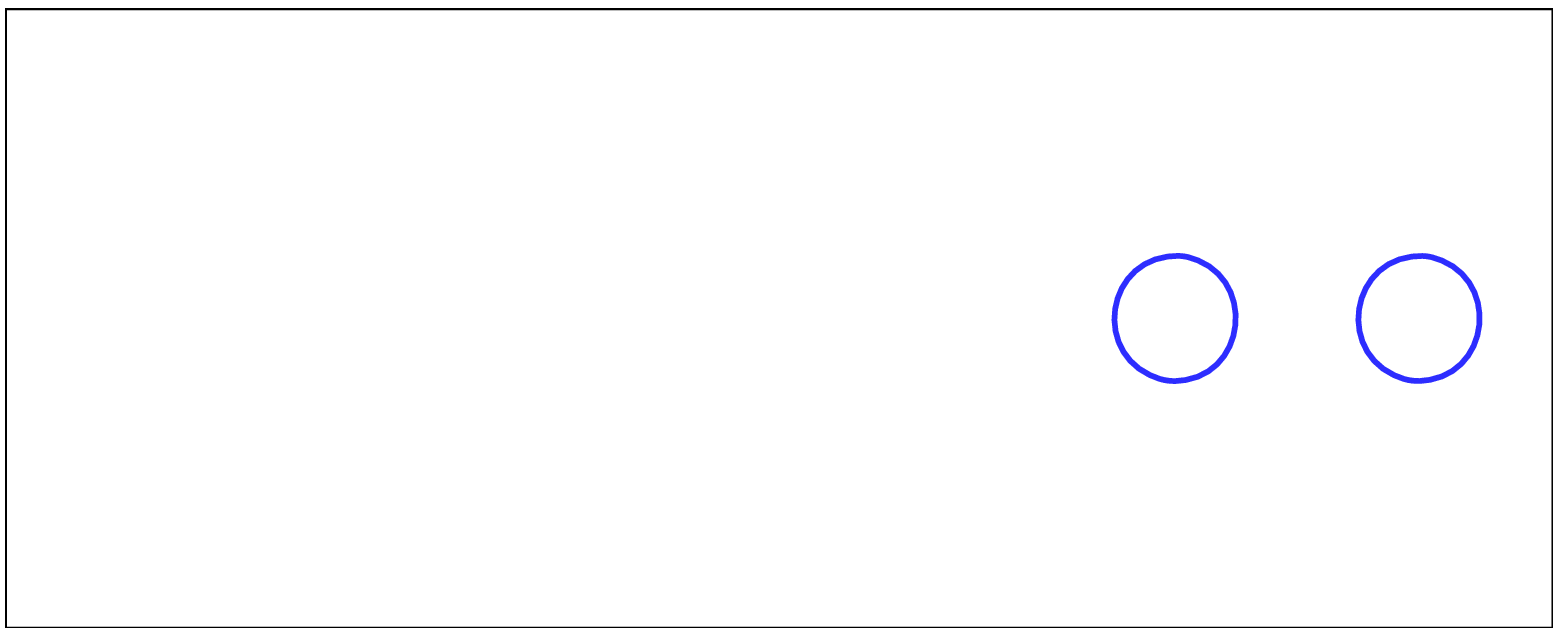}
	\end{minipage}
	\begin{minipage}[c]{0.49\textwidth}
		\includegraphics[width=\textwidth]{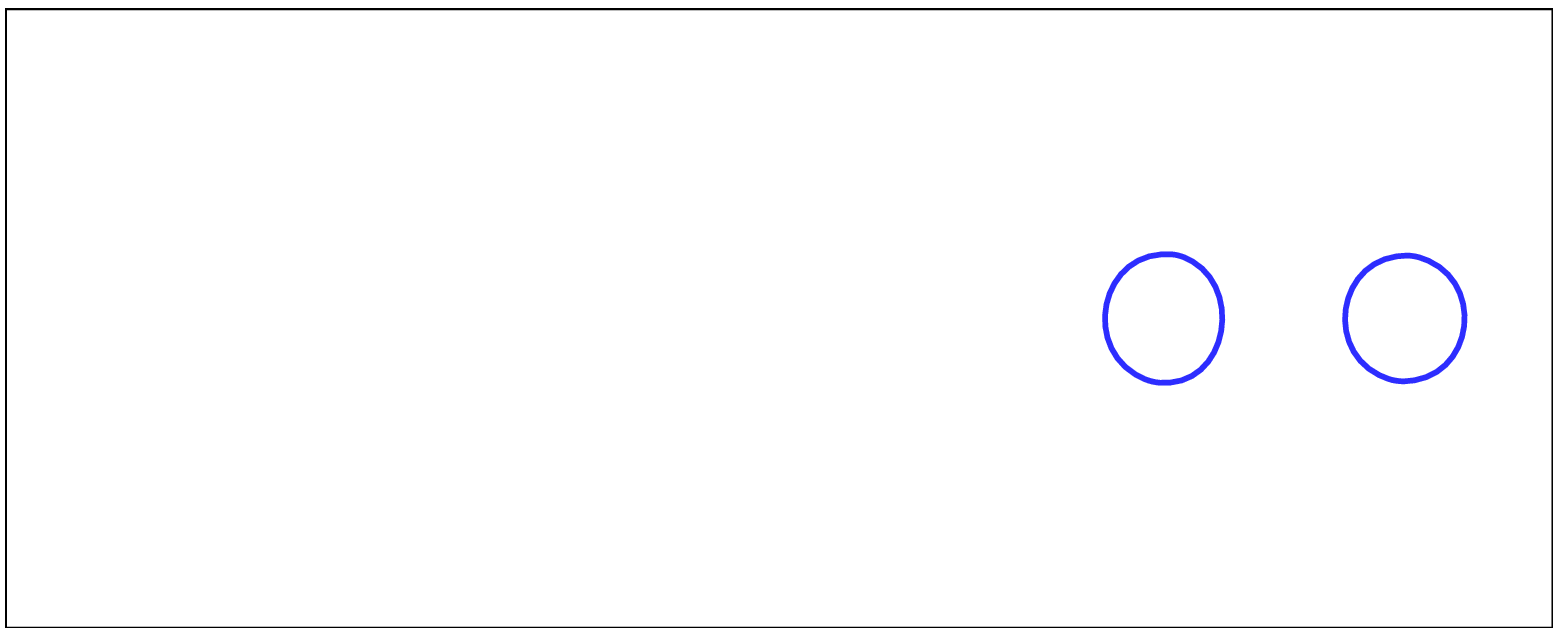}
	\end{minipage}

	\begin{minipage}[c]{0.49\textwidth}
		\includegraphics[width=\textwidth]{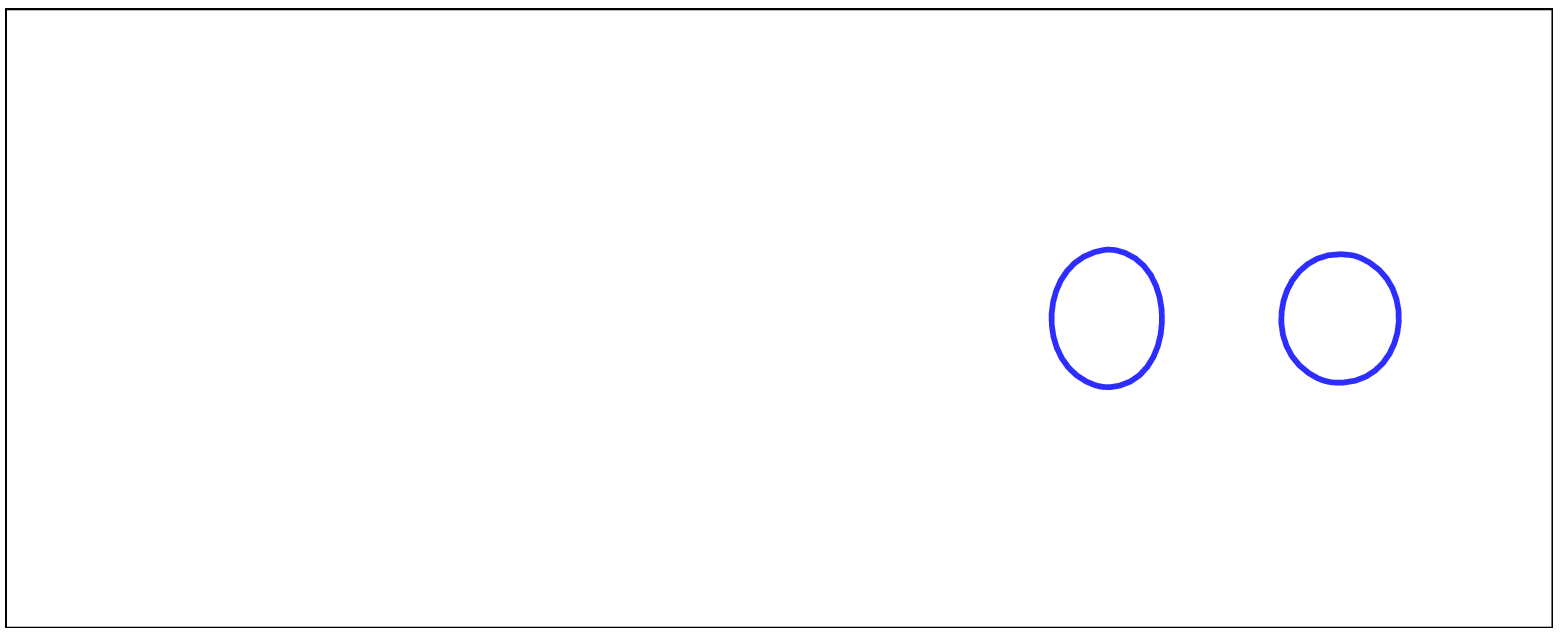}
	\end{minipage}
	\begin{minipage}[c]{0.49\textwidth}
		\includegraphics[width=\textwidth]{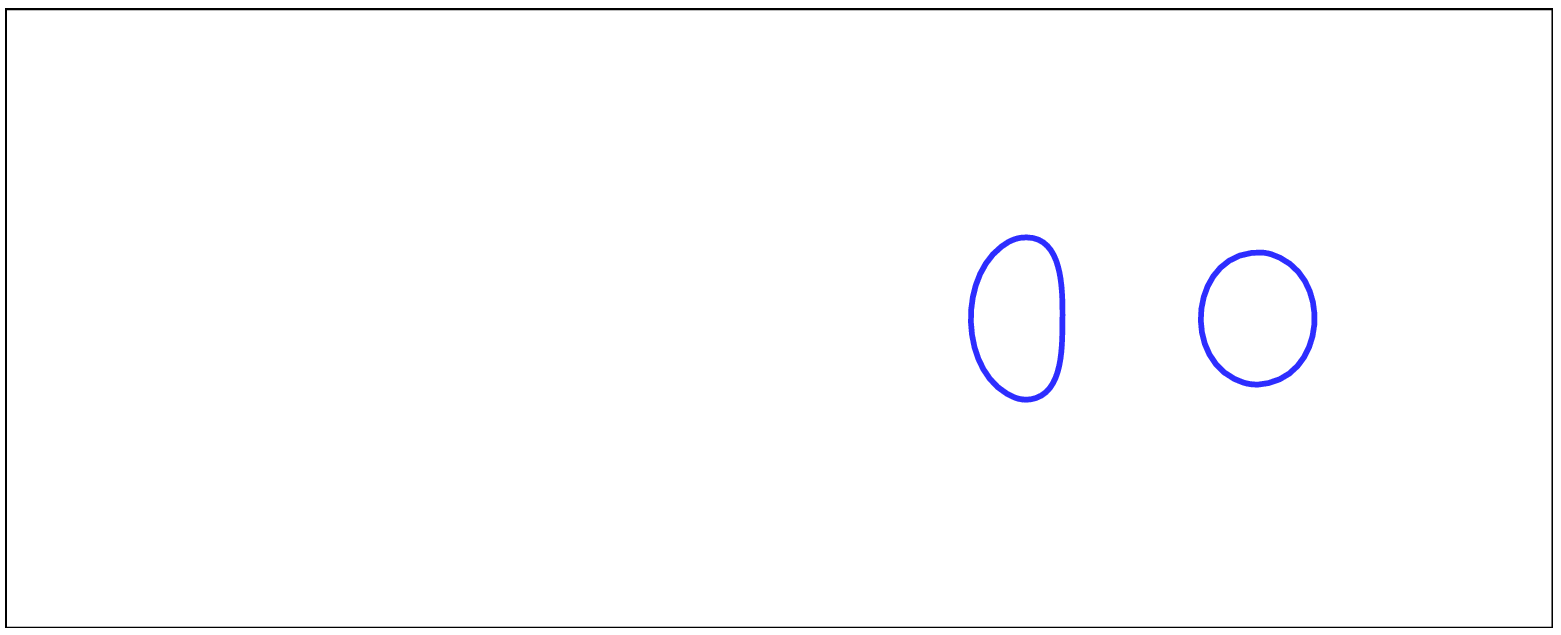}
	\end{minipage}
	\begin{minipage}[c]{0.49\textwidth}
		\includegraphics[width=\textwidth]{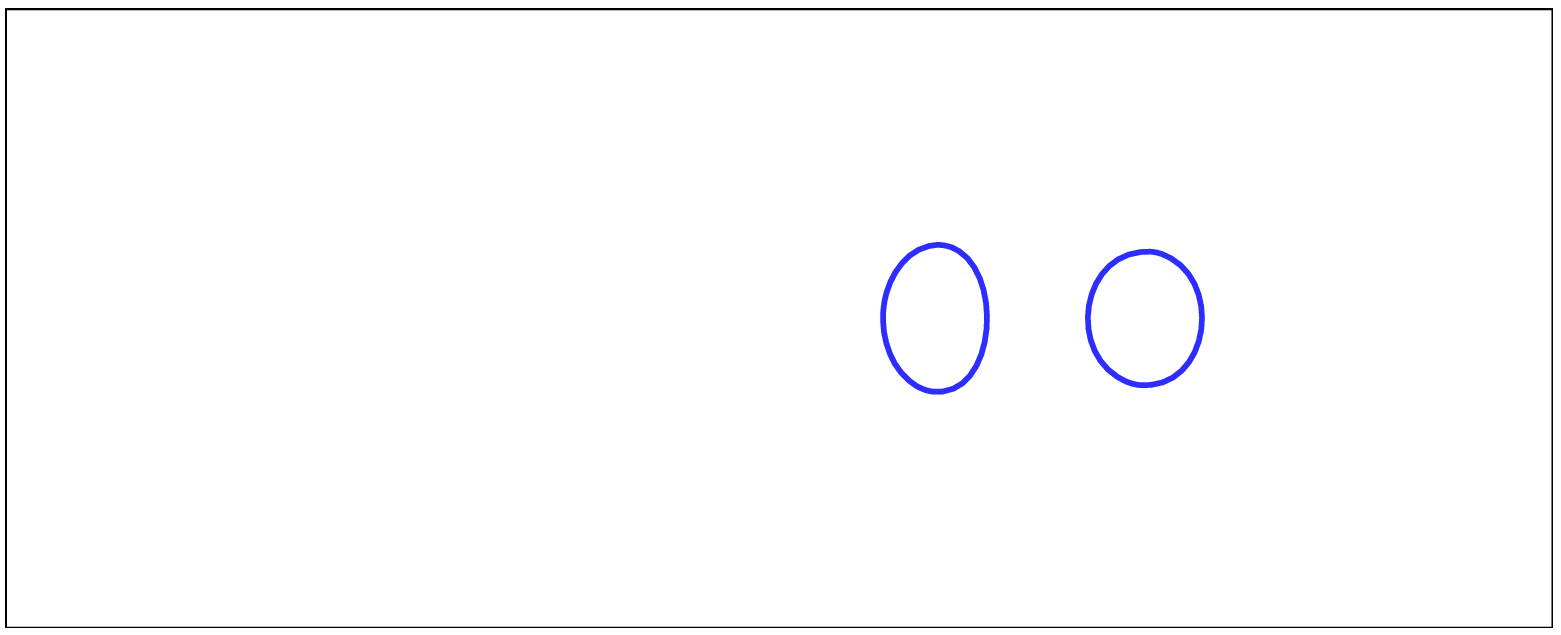}
	\end{minipage}
	\begin{minipage}[c]{0.49\textwidth}
	\includegraphics[width=\textwidth]{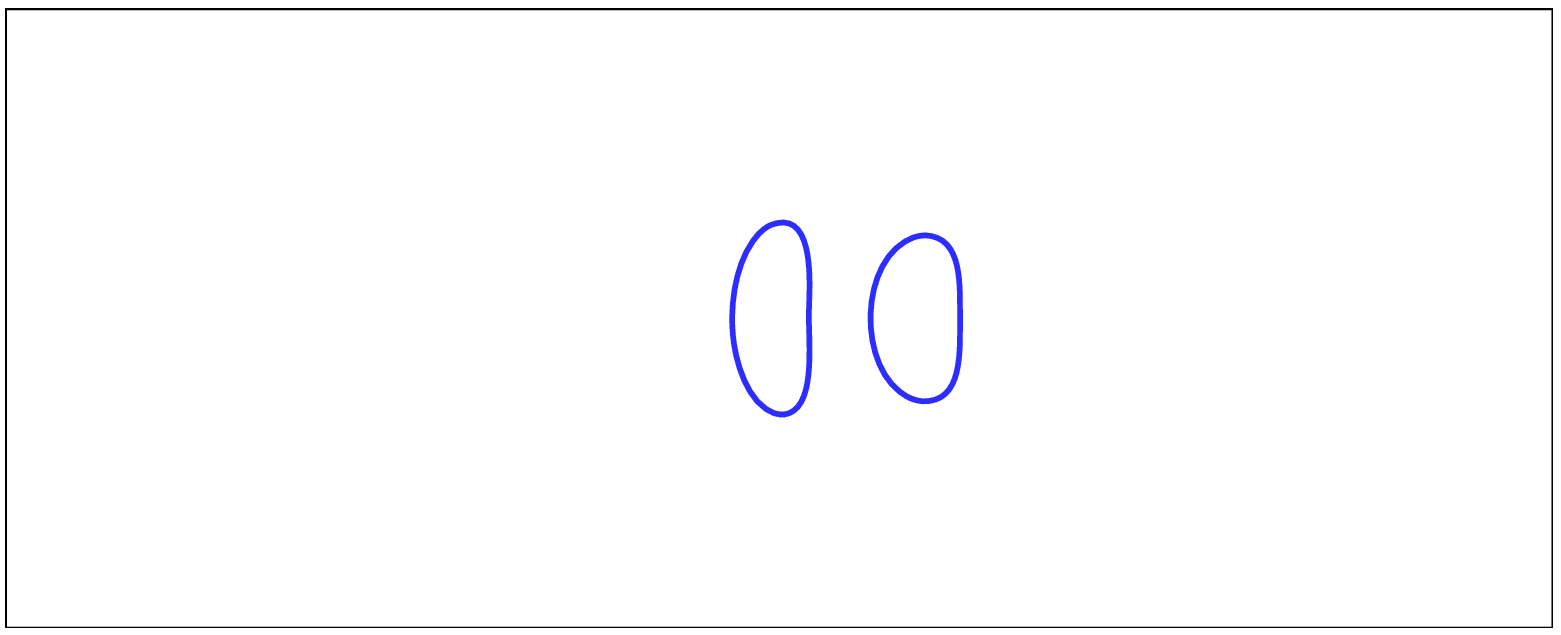}
    \end{minipage}
	\begin{minipage}[c]{0.49\textwidth}
	\includegraphics[width=\textwidth]{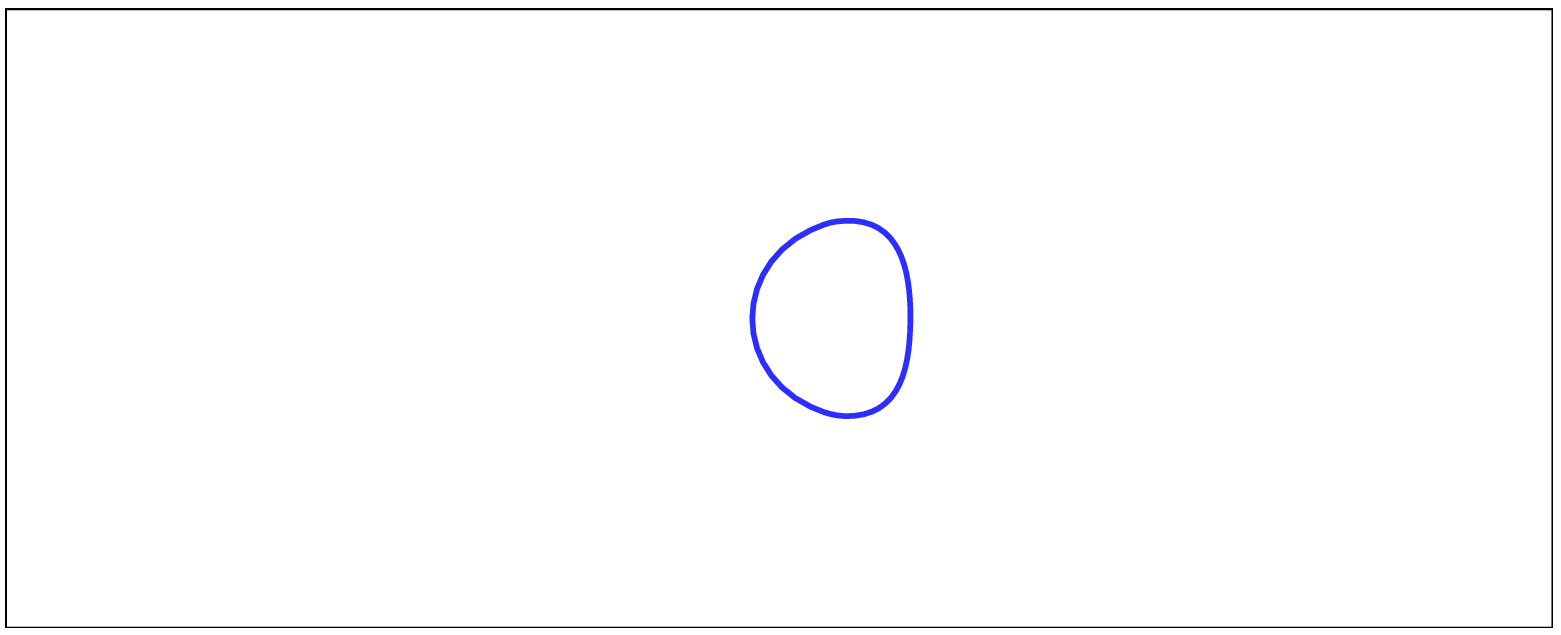}
\end{minipage}
	\begin{minipage}[c]{0.49\textwidth}
	\includegraphics[width=\textwidth]{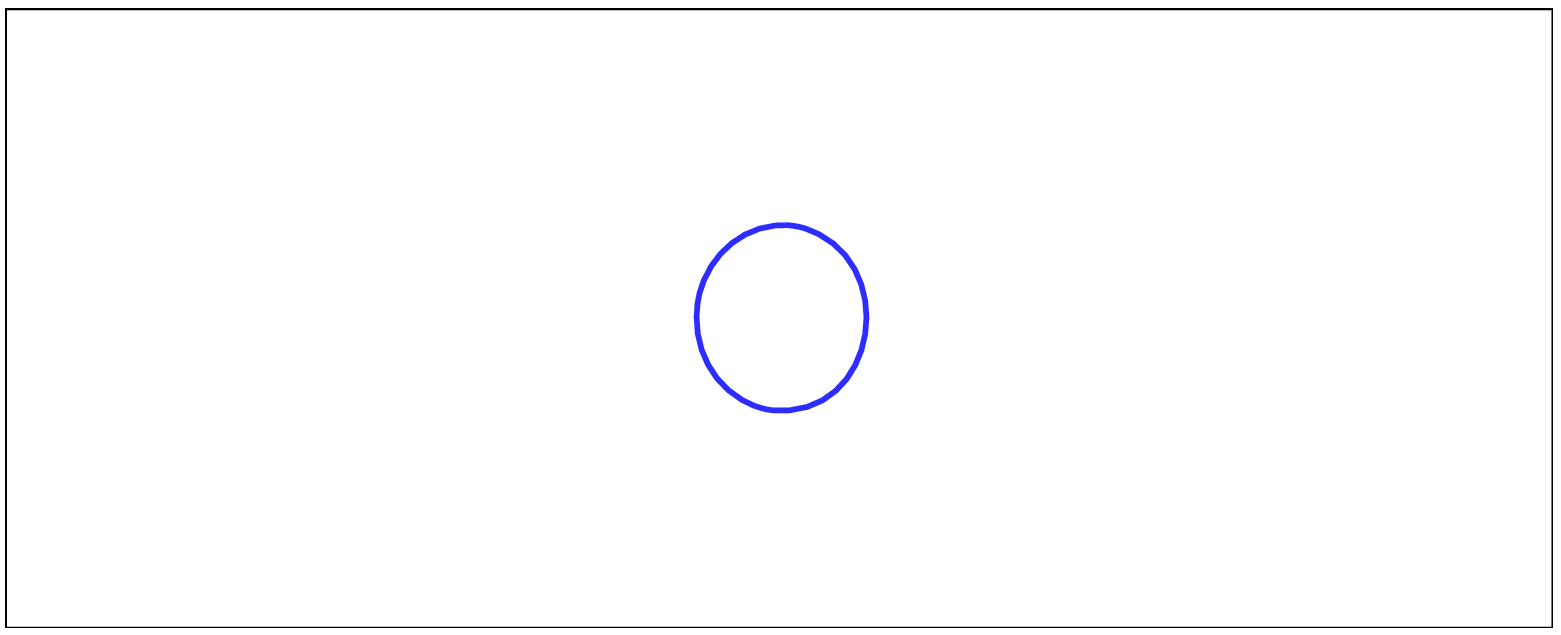}
\end{minipage}
	\caption{Snapshots of thermocapillary flow of two recalcitrant bubbles in a mircochannel for  ${\sigma _{TT}}=0.2$ (left) and  ${\sigma _{TT}}=0.6$ (right). The time instants are normalized by $L/u_{ave}$, and from the top to bottom, their values are taken as $t^*=0.00125$,  $t^*=0.00625$,  $t^*=0.01875$ and $t^*=0.04375$, respectively.}
	\label{figw10}
\end{figure}

\begin{figure}[H]
	\centering
	\subfigure[]{\label{figw11a}
		\includegraphics[width=0.48\textwidth]{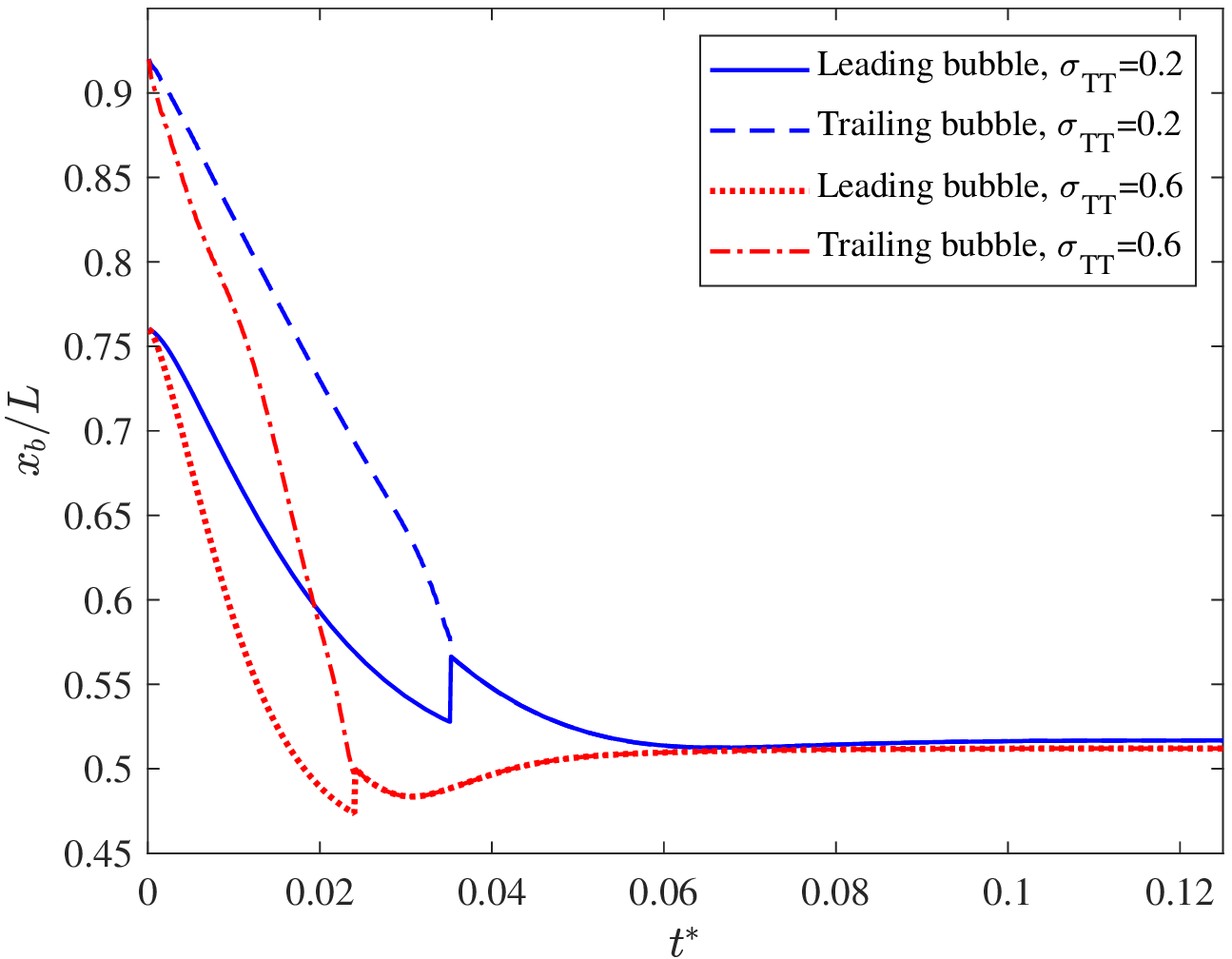}
	}
	\subfigure[]{\label{figw11b}
		\includegraphics[width=0.48\textwidth]{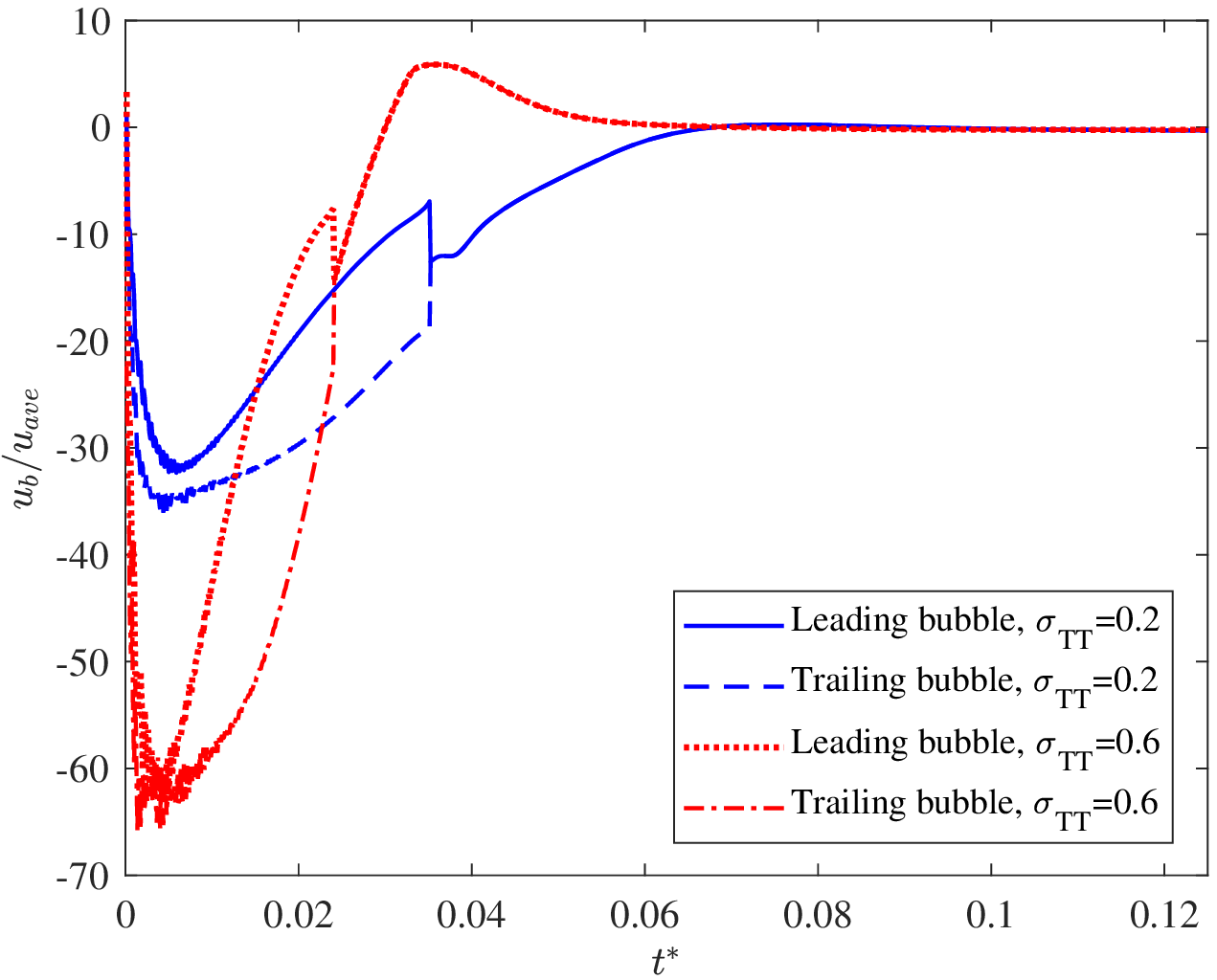}
	}
	\caption{The time evolution of bubbles position (a) and their velocity (b) for different  ${\sigma _{TT}}$.}
	
	\label{figw11}
\end{figure}

Fig. \ref{figw10} depicts four typical scenic representations of the two recalcitrant bubbles at two different  ${\sigma _{TT}}$, where the normalized time is given by $t^*=tu_{ave}/L$ with $t$ representing the iteration step. As it is apparent, the bubbles in both  cases are able to move against the flow direction under the effects of the thermocapillary force, proving that the present method is able to reproduce the weird behavior of the recalcitrant bubble in a thermocapillary flow. As shown in Fig. \ref{figw10}, apart from the deformation process observed in the case of one single recalcitrant bubble, the leading and trailing bubble also encounter the interaction and the coalescence before reaching the equilibrium point. On the other hand, with the fluid flow from left to right, the drag force by the viscous flow has a profound effect on the leading bubble, and causing it to experience more deformation in contrast to the trailing bubble. As for the effect of ${\sigma _{TT}}$, it is noted that the as ${\sigma _{TT}}$ increases, the bubbles deform to a larger degree and reach the equilibrium point with a relatively smaller time consuming. In fact, it is known that the movement of the recalcitrant bubbles is the result of the interaction between the thermocapillary force (driving force) and the drag force (induced by fluid flow). Since the fluid velocity is the same for both cases, the drag force for these two cases are the same. In such a case, the difference of the droplet dynamics is caused by the variation of the thermocapillary force. Recalling the definition of surface tension [i.e, Eq. (\ref{eq_63})], it is obvious that the gradient of the surface tension is increased with increasing ${\sigma _{TT}}$, so does the deformation of the bubbles. Also, the increased thermocapillary force in turn results in the bubbles to move with a higher velocity. To quantitative investigate migration velocity and the local position of each bubble, Fig. \ref{figw11} shows the time variation of the  numerically predicted results by the present LB scheme. At the initial stage of migration, the bubble location is decreased rapidly together with an increase of the migration velocity. In particularly, for a higher ${\sigma _{TT}}$, the above phenomenon becomes more remarkable, which is in line with the previous discussions. After that, the bubbles combined to form a larger bubble coupled with a relatively smaller velocity, and then move towards the upstream until it reaches the equilibrium position. It is note worthy to mention that although the values of ${\sigma _{TT}}$ are not the same in these two cases, the final bubble location are only slightly different in theses two cases.   
    
\subsection{Conclusion}
Thermocapillary flows play an important role in various industrial applications, but numerical modeling of such flows with thermophysical parameters contrast still remains a challenging task in the LB community. In this article, we present a phase-field based LB scheme for simulating thermocapillary flows, which is able to deal with thermophysical parameters contrast. To capture the interface of the two-phase flows, the conservative Allen-Cahn and Navier-Stokes equations are solved by using two different distribution functions. Besides, in order to solve the temperature equation within the framework of the LB approach, the evolution equation is carefully designed, which recover the energy equation correctly through the Chapman-Enskog analysis. Different from previous LB models where the space derivatives of the heat capacitance must be solved with an finite difference scheme, the present LB model could avoid the above issue, which makes the present LB approach retain the main merits of the LB method. The method is first validated by  thermocapillary flows with two superimposed planar fluids and thermocapillary migration of deformable droplets. In the former test, it is found that the numerical results predicted by the present method are in good agreement with the analytical solutions. In addition, it is reported that the horizontal and vertical velocities are smooth even near 
the areas of phase interface, indicating our scheme is more stable compared with Yue et al.'s model. In the latter case, we simulated thermocapillary flows with density and thermocapillary parameters contrasts, and also conduced comparisons with the theoretic solutions given by Young et al.. Good consistence can be seen again.
 Further, we also investigated the migration behavior for a range of Marangoni numbers, and the numerical results are in line with the previous studies. Last, to show the present method's capability, we simulated the thermocapillary flow of two recalcitrant bubbles, and investigated the influence of the  ${\sigma _{TT}}$ (the second derivative of the surface tension with respect to the temperature), where the surface tension is set to be a parabolic function of the temperature. It is found that the leading and trailing bubbles are both move against the flow direction, and the leading bubble deform to a larger degree in contrast the trailing one. As ${\sigma _{TT}}$ increases, the migration velocity also increases owing to the enhancement of the thermocapillary force. Thus, the total time consumed to reach the equilibrium position decreases.  In conclusion, our numerical results demonstrate the feasibility of the proposed LB method, and the merits of the present scheme show that it is a promising candidate for simulating thermocapillary flows with complex boundaries.

\subsection{Acknowledgments}
This work is financially supported by the National Natural Science Foundation of China (Grant No. 12002320) and the Fundamental Research Funds for the Central Universities (Grant No. CUGQT2023001).


\begin{thebibliography}{1}
\bibitem{striven1960}L. E. Strive, Dynamics of a fluid interface equation of motion for Newtonian surface fluids, Chem. Eng. Sci. 12, 98 (1960).

\bibitem{scriven1960nature}L. E. Scriven and C . V. Sternling, The marangoni effects, Nature 187, 186 (1960).

\bibitem{shen1990jfm}Y. Shen, G. P. Neitzel, D. F. Jankowski and H. D. Mittelmann, Energy stability of thermocapillary convection in a model of the float-zone crystal-growth process, J. Fluid Mech. 217, 639 (1990).

\bibitem{mills1998marabgo}K. C. Mills, B. J. Keene, R. F. Brooks and A. Shirali, Marangoni effects in welding, Philos. Trans. R. Soc. A-Math. Phys. Eng. Sci. 356, 911 (1998).


\bibitem{jasnow1996coarse}P. Dell’Aversana, J. R.  Banavar, J. Koplik, Suppression of coalescence by shear and temperature gradients, Phys. Fluids  9, 15 (1996). 

\bibitem{young1959the}N. O. Young, J. S. Goldstein and M. J. Block, The motion of bubbles in a vertical temperature gradient, J. Fluid Mech. 6, 350 (1959).

\bibitem{schatz2001experiment}M. F. Schatz and G. P. Neitzel, Experiments on thermocapillary instabilities, Annu. Rev. Fluid Mech. 33, 93 (2001).

\bibitem{robert2012thermal}M. R. de Saint Vincent and J.-P. Delville, Thermocapillary migration in small-scale temperature gradients: Application to optofluidic drop dispensing, Phys. Rev. E 85, 026310 (2012).

\bibitem{kamotani2000micro}Y. Kamotani, S. Ostrach and J. Masud, Microgravity experiments and analysis of oscillatory thermocapillary flows in cylindrical containers, J. Fluid Mech. 410, 211 (2000).

\bibitem{kang2019the}Q. Kang, J. Wang, L. Duan, Y. Su, J. He, D. Wu and W. Hu, The volume ratio effect on flow patterns and transition processes of thermocapillary convection, J. Fluid Mech. 868, 560 (2019).

\bibitem{karbalaei2016thermo}A. Karbalaei, R. Kumar and H. J. Cho, Thermocapillarity in microfluidics—A review, Micromachines 7, 13 (2016).

\bibitem{nas2033therm}S. Nas and G. Tryggvason, Thermocapillary interaction of two bubbles or drops, Int. J. Multiph. Flow 29, 1117 (2003).


\bibitem{ma2013numerical}C. Ma and D. Bothe, Numerical modeling of thermocapillary two-phase flows with evaporation using a two-scalar approach for heat transfer, J. Comput. Phys. 233, 552 (2013).

\bibitem{zhao2011topo}J. Zhao, L. Zhang, Z. Li and W. Qin, Topological structure evolvement of flow and temperature fields in deformable drop Marangoni migration in microgravity, Int. J. Heat Mass Transf. 54, 4655 (2011).

\bibitem{shyy1996comput}W. Shyy, R. W. Smith, H. S. Udaykumar and M. M. Rao, Computational Fluid Dynamics with Moving Boundaries (Taylor \& Francis, London, 1996).

\bibitem{kruge2017the} T. Krüger, H. Kusumaatmaja, A. Kuzmin, O. Shardt, G. Silva, and E. M. Viggen, The Lattice Boltzmann Method: Principles and Practice (Springer, Cham, Switzerland, 2017).

\bibitem{guobook2013lattice}Z. Guo and C. Shu, Lattice Boltzmann Method and Its Applications in Engineering (World Scientific Publishing Co. Pte. Ltd.,Singapore, 2013).


\bibitem{huang2015multiphase}H. Huang, M. Sukop, X.-Y. Lu, Multiphase lattice boltzmann methods: theory and application (John Wiley \& Sons; Oxford, 2015).

\bibitem{lireview2016lattice}Q. Li, K. H. Luo, Q. J. Kang, Y. L. He, Q. Chen and Q. Liu, Lattice Boltzmann methods for multiphase flow and phase-change heat transfer, Prog. Energy Combust. Sci. 52, 62 (2016).

\bibitem{huang2022three}J. Huang, L. Wang and K. He, Three-dimensional study of double droplets impact on a wettability-patterned surface, Comput. Fluids 248, 105669 (2022).

\bibitem{wang2022thermal}L. Wang, J. Huang and K. He, Thermal lattice Boltzmann model for liquid-vapor phase change, Phys. Rev. E 106, 055308 (2022).

\bibitem{gupta2016lattice}A. Gupta, M. Sbragaglia, D. Belardinelli and K. Sugiyama, Lattice Boltzmann simulations of droplet formation in confined channels with thermocapillary flows, Phys. Rev. E 94, 063302 (2016).

\bibitem{liu2012modeling}H. Liu, Y. Zhang and A. J. Valocchi, Modeling and simulation of thermocapillary flows using lattice Boltzmann method, J. Comput. Phys. 231, 4433 (2012).

\bibitem{liujcp2015modelling}H. Liu and Y. Zhang, Modelling thermocapillary migration of a microfluidic droplet on a solid surface, J. Comput. Phys. 280, 37 (2015).

\bibitem{liu2017alattice}H. Liu, L. Wu, Y. Ba Y and G. Xi, A lattice Boltzmann method for axisymmetric thermocapillary flows, Int. J. Heat Mass Transf. 104, 337 (2017).

\bibitem{liujcp2014lattice}H. Liu, A. J.  Valocchi, Y. Zhang and Q. Kang, Lattice Boltzmann phase-field modeling of thermocapillary flows in a confined microchannel, J. Comput. Phys. 256, 334 (2014).

\bibitem{zheng2016continuous}L. Zheng, S. Zheng and Q. Zhai, Continuous surface force based lattice Boltzmann equation method for simulating thermocapillary flow, Phys. Lett. A 380, 596 (2016).

\bibitem{liupre2013phase}H. Liu, A. J.  Valocchi, Y. Zhang and Q. Kang, Phase-field-based lattice Boltzmann finite-difference model for simulating thermocapillary flows, Phys. Rev. E 87, 013010 (2013).

\bibitem{leejcp2010lattice}T. Lee and L. Liu, Lattice Boltzmann simulations of micron-scale drop impact on dry surfaces, J. Comput. Phys 229, 8045 (2010).

\bibitem{lou2012effects}Q. Lou, Z. L. Guo, and B. C. Shi, Effects of force discretization on mass conservation in lattice Boltzmann equation for two-phase flows, Europhys. Lett. 99, 64005 (2012).

\bibitem{qiaomodel2018ate}L. Qiao, Z. Zeng, H. Xie, L. Zhang, L. Wang and Y. Lu, Modeling thermocapillary migration of interfacial droplets by a hybrid lattice Boltzmann finite difference scheme, Appl. Therm. Eng. 131, 910 (2018).

\bibitem{majidi2020single}M. Majidi, H.-H.-A. Reza and M. H. Rahimian, Single recalcitrant bubble simulation using a hybrid lattice Boltzmann finite difference model, Int. J. Multiph. Flow 127, 103289 (2020).

\bibitem{mitchell2021compu}T. R. Mitchell, M. Majidi, M. H. Rahimian, Computational modeling of three-dimensional thermocapillary flow of recalcitrant bubbles using a coupled lattice Boltzmann-finite difference method, Phys Fluids 33, 032108 (2021).

\bibitem{wang2019ammalatt}L. Wang, Y. Zhao, X. Yang, B. Shi and Z. Chai, A lattice Boltzmann analysis of the conjugate natural convection in a square enclosure with a circular cylinder, Appl. Math. Model. 71, 31 (2019).

\bibitem{guo2002acoupled}Z. Guo, B. Shi and C. Zheng, A coupled lattice BGK model for the Boussinesq equations, Int. J. Numer. Methods Fluids 39, 325 (2002).

\bibitem{hu2019adiffuse}Y. Hu, D. Li, X. Niu and S. Shu, A diffuse interface lattice Boltzmann model for thermocapillary flows with large density ratio and thermophysical parameters contrasts, Int. J. Heat Mass Transf. 138, 809 (2019).

\bibitem{yue2022improved}L. Yue, Z. Chai and H. Wang, Improved phase-field-based lattice Boltzmann method for thermocapillary flow, Phys. Rev. E 105, 015314 (2022).

\bibitem{wang2016comparative}H. Wang, Z. Chai and B. Shi, Comparative study of the lattice Boltzmann models for Allen-Cahn and Cahn-Hilliard equations, Phys. Rev. E 94, 033304 (2016).

\bibitem{liang2018phase}H. Liang, J. Xu, J. Chen, H. Wang, Z. Chai and B. Shi, Phase-field-based lattice Boltzmann modeling of large-density-ratio two-phase flows, Phys. Rev. E 97, 033309 (2018).

\bibitem{wang2019abrief}H. Wang, X. Yuan, H. Liang, Z. Chai and B. Shi, A brief review of the phase-field-based lattice Boltzmann method for multiphase flows, Capillarity 2, 33 (2019).

\bibitem{sun2007jcpsharp} Y. Sun and C. Beckermann, Sharp interface tracking using the phase-field equation, J. Comput. Phys. 220, 626 (2007).

\bibitem{folch1999phase}R. Folch and J. Casademunt, A. Hernández-Machado. Phase-field model for Hele-Shaw flows with arbitrary viscosity contrast. I. Theoretical approach, Phys. Rev. E 60, 1724 (1999).

\bibitem{unverdi1992afront} S. O. Unverdi and G. Tryggvason, A front-tracking method for viscous, incompressible, multi-fluid flows, J. Comput. Phys. 100, 25 (1992).

\bibitem{ma2011direct}C. Ma and D. Bothe, Direct numerical simulation of thermocapillary flow based on the volume of fluid method, Int. J. Multiph. Flow 37, 1045 (2011).

\bibitem{ma2013jcpnumeri}C. Ma and D. Bothe, Numerical modeling of thermocapillary two-phase flows with evaporation using a two-scalar approach for heat transfer, J. Comput. Phys. 233, 552 (2013).

\bibitem{qian1992lattice} Y. H. Qian, D. d’Humires and P. Lallemand, Lattice BGK models for Navier-Stokes equation, Europhys. Lett. 17, 479 (1992).

\bibitem{ginzburg2008two} I. Ginzburg, F. Verhaeghe and D. d’Humières, Two-relaxation-time lattice Boltzmann scheme: About parametrization, velocity, pressure and mixed boundary conditions, Commun. Comput. Phys. 3, 427 (2008).

\bibitem{lallemand2000theory} P. Lallemand and L. S. Luo, Theory of the lattice Boltzmann method: Dispersion, dissipation, isotropy, Galilean invariance,and stability, Phys. Rev. E 61, 6546 (2000).

\bibitem{cartalade2016lattice}A. Cartalade, A. Younsi and M. Plapp, Lattice Boltzmann simulations of 3D crystal growth: Numerical schemes for a phase-field model with anti-trapping current, Comput. Math. Appl. 71, 1784 (2016).

\bibitem{sun2019ananiso}D. Sun, H. Xing and X. Dong, An anisotropic lattice Boltzmann-Phase field scheme for numerical simulations of dendritic growth with melt convection, Int. J. Heat Mass Transf. 133, 1240 (2019).

\bibitem{chai2013pre}Z. Chai and T. S. Zhao, Lattice Boltzmann model for the convection-diffusion equation, Phys. Rev. E 87, 063309 (2013).

\bibitem{pendse2010ananalytic}B. Pendse and A. Esmaeeli, An analytical solution for thermocapillary-driven convection of superimposed fluids at zero Reynolds and Marangoni numbers, Int. J. Therm. Sci. 49, 1147 (2010).

\bibitem{balasubramaniam2000the}R. Balasubramaniam and R. S. Subramanian, The migration of a drop in a uniform temperature gradient at large Marangoni numbers, Phys. Fluids 12, 733 (2000).

\bibitem{xie2005experiment}J. C. Xie, H. Lin and P. Zhang, Experimental investigation on thermocapillary drop migration at large Marangoni number in reduced gravity, J. Colloid Interface Sci. 285, 737 (2005).

\bibitem{shankar1988the}N. Shankar and R. S. Subramanian, The Stokes motion of a gas bubble due to interfacial tension gradients at low to moderate Marangoni numbers,  J. Colloid Interface Sci. 123, 512 (1988).

\bibitem{yin2008thermo}Z. Yin, P. Gao and W. Hu, Thermocapillary migration of nondeformable drops, Phys. Fluids 20, 082101 (2008).

\bibitem{shanahan2014reca}M. E. R. Shanahan and K. Sefiane, Recalcitrant bubbles, Sci. Rep. 4, 4727 (2014).

\bibitem{zou1997on} Q. Zou and X. He, On pressure and velocity boundary conditions for the lattice Boltzmann BGK model, Phys. Fluids 9, 1591 (1997).
\end{thebibliography}
\end{document}